\documentclass[useAMS,usenatbib]{mn2e}

\usepackage{epsfig,color,pifont}

\usepackage{natbib}
\bibpunct{(}{)}{;}{a}{}{,}
\def\spose#1{\hbox to 0pt{#1\hss}}
\def\ltsimm{\mathrel{\spose{\lower 3pt\hbox{$\sim$}}
        \raise 2.0pt\hbox{$<$}}}
\def\gtsimm{\mathrel{\spose{\lower 3pt\hbox{$\sim$}}
        \raise 2.0pt\hbox{$>$}}}

\def\km{{\rm\thinspace km}}
\def\cm{{\rm\thinspace cm}}

\def\s{{\rm\thinspace s}}
\def\yr{{\rm\thinspace yr}}
\def\g{{\rm\thinspace g}}
\def\kmps{\hbox{${\rm\km\s^{-1}\,}$}}

\def\erg{{\rm\thinspace erg}}
\def\Hz{{\rm\thinspace Hz}}

\def\ster{{\rm\thinspace ster}}
\def\ergps{\hbox{${\rm\erg\s^{-1}\,}$}}
\def\Rsol{\hbox{${\rm\thinspace R_{\odot}}$}}
\def\Msol{\hbox{${\rm\thinspace M_{\odot}}$}}
\def\Lsol{\hbox{${\rm\thinspace L_{\odot}}$}}
\def\Msolpyr{\hbox{${\rm\Msol\yr^{-1}\,}$}}
\def\pcm{\hbox{${\rm\cm^{-1}\,}$}}
\def\pcm2{\hbox{${\rm\cm^{-2}\,}$}}
\def\pcm3{\hbox{${\rm\cm^{-3}\,}$}}
\def\ergpscm3Hz{\hbox{${\rm\ergps\cm^{-3}\Hz^{-1}\,}$}}
\def\ergpscm3Hzster{\hbox{${\rm\ergps\cm^{-3}\Hz^{-1}\ster^{-1}\,}$}}
\def\gpcm3{\hbox{${\rm\g\cm^{-3}\,}$}}
\def\ergpcm2{\hbox{${\rm\erg\cm^{-2}\,}$}}
\def\ergpcm3{\hbox{${\rm\erg\cm^{-3}\,}$}}
\def\phpscm2{\hbox{${\rm photons\s^{-1}\cm^{-2}\,}$}}

\def\etacar{$\eta\thinspace\rm{Car}\;$}

\title[3D modelling of the colliding winds in
$\eta\thinspace$Carinae]{3D modelling of the colliding winds in
$\eta\thinspace$Carinae - evidence for radiative inhibition}

\author[E.~R.~Parkin, J.~M.~Pittard, M.~F.~Corcoran, K.~Hamaguchi, and
I.~R.~Stevens] {E. R. Parkin$^{1}$\thanks{E-mail:
phy1erp@leeds.ac.uk}, J.~M.~Pittard$^{1}$, M.~F.~Corcoran$^{2,3}$,
K.~Hamaguchi$^{2,4}$, and \newauthor I.~R.~Stevens$^{5}$ \\
$^{1}$School of Physics and Astronomy, The University of Leeds,
Woodhouse Lane, Leeds LS2 9JT, UK \\ $^{2}$CRESST and X-ray
Astrophysics Laboratory, NASA/GSFC, Greenbelt, MD 20771, USA\\
$^{3}$Universities Space Research Association, 10211 Wisconsin Circle,
Suite 500 Columbia, MD 21044, USA\\ $^{4}$Department of Physics,
Universtiy of Maryland, Baltimore County, 1000 Hilltop Circle,
Baltimore, MD 21250, USA\\ $^{5}$School of Physics and Astronomy,
University of Birmingham, Edgbaston, Birmingham B15 2TT, UK}

\begin{document}

\date{Accepted ... Received ...; in original form ...}

\pagerange{\pageref{firstpage}--\pageref{lastpage}} \pubyear{2008}

\maketitle

\label{firstpage}

\begin{abstract}
The X-ray emission from the super-massive star \etacar is simulated
using a three dimensional model of the wind-wind collision. In the
model the intrinsic X-ray emission is spatially extended and energy
dependent. Absorption due to the unshocked stellar winds and the
cooled postshock material from the primary LBV star is calculated as
the intrinsic emission is ray-traced along multiple sightlines through
the 3D spiral structure of the circumstellar environment. The
observable emission is then compared to available X-ray data,
including the lightcurve observed by the \textit{Rossi X-ray Timing
Explorer (RXTE)} and spectra observed by \textit{XMM-Newton}. The
orientation and eccentricity of the orbit are explored, as are the
wind parameters of the stars and the nature and physics of their close
approach. Our modelling supports a viewing angle with an inclination
of $\simeq 42^{\circ}$, consistent with the polar axis of the
Homunculus nebula \citep{Smith:2006}, and the projection of the
observer's line-of-sight onto the orbital plane has an angle of
$\simeq 0 - 30^{\circ}$ in the prograde direction on the apastron side
of the semi-major axis.

However, there are significant discrepancies between the observed and
model lightcurves and spectra through the X-ray minimum. In
particular, the hard flux in our synthetic spectra is an order of
magnitude greater than observed. This suggests that the hard X-ray
emission near the apex of the wind-wind collision region (WCR)
`switches off' from periastron until 2 months afterwards. Further
calculations reveal that radiative inhibition significantly reduces
the preshock velocity of the companion wind. As a consequence the hard
X-ray emission is quenched, but it is unclear whether the long
duration of the minimum is due solely to this mechanism alone. For
instance, it is possible that the collapse of the WCR onto the surface
of the companion star, which would be aided by significant inhibition
of the companion wind, could cause an extended minimum as the
companion wind struggles to re-establish itself as the stars
recede. For orbital eccentricities, $e \ltsimm 0.95$, radiative
braking prevents a wind collision with the companion star's
surface. Models incorporating a collapse/disruption of the WCR and/or
reduced preshock companion wind velocities bring the predicted
emission and the observations into much better agreement.

\end{abstract}


\section{Introduction}
\label{sec:intro}
The super-massive star \etacar provides a unique laboratory for the
study of massive star evolution. Situated in the Carina nebula, at a
distance of 2.3 kpc \citep{Davidson:1997}, \etacar has been identified
as a luminous blue variable star (LBV), a short-lived ($\sim 10^{4}\rm
{yr}$) phase in the life of a massive star occuring after the main
sequence and preceding the Wolf-Rayet phase. \etacar is believed to
have a current mass of $\simeq 80-120\Msol$, and an initial mass
$\gtsimm 150\Msol $ \citep{Hillier:2001}. Over the past 200 years
episodes of rapid mass loss have been observed from the star, and in
1843 the ``Great Eruption'' ejected more than 10 \Msol\thinspace of
matter \citep{Smith:2003a} from the central object over a $\sim$ 20
year period, forming the bipolar nebula known as the Homunculus. In
the 1890's a further smaller eruption with a mass loss of $\sim 1
\Msol$ produced the Little Homunculus \citep{Ishibashi:2003}.  The
central object is now shrouded by a large amount of dense absorbing
gas and hot and cold dust which makes observations at optical and UV
wavelengths difficult.

\begin{figure*}
\begin{center}
\psfig{figure=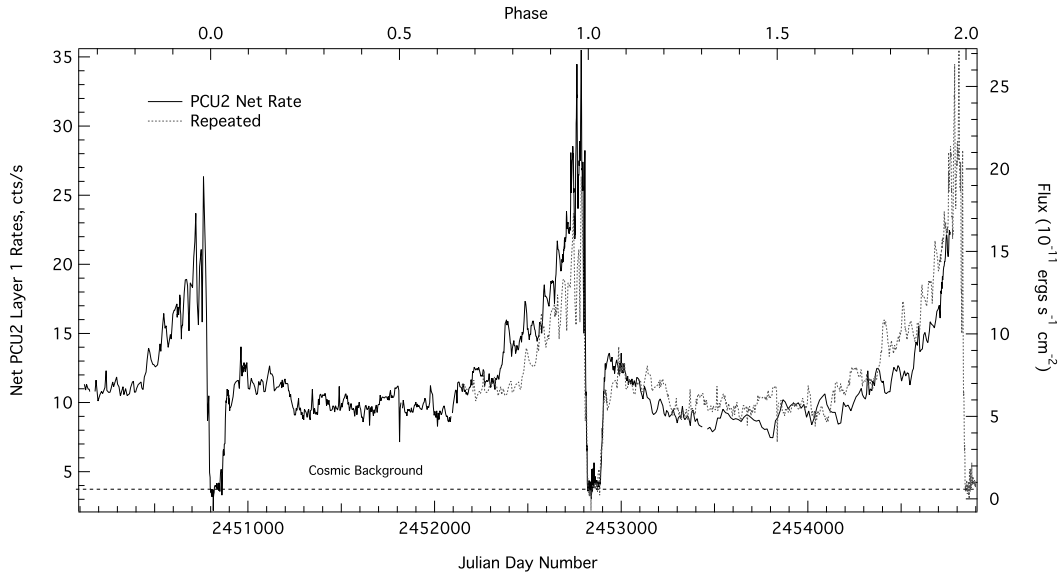,width=14.0cm}
\caption[]{The 2-10 keV X-ray lightcurve of \etacar corrected for
instrumental background over two periastron passages taken with the
\textit{RXTE} satellite \citep{Corcoran:2005}. The data from $\sim
1996-2003.5$ (dotted) has been super-imposed over the 2002-present
data for comparison. For the majority of the orbit the flux is broadly
constant. Prior to periastron passage there is a distinctive rise to
maximum, then a sharp fall to a deep minimum. There is a noticeable
difference between the peak magnitudes prior to the minimum of the
last two cycles. The dotted horizontal line is an estimate of the flux
due to cosmic background which contaminates the \textit{RXTE/PCU2}
field of view.}
\label{fig:rxtecurve}
\end{center}
\end{figure*}

Although many questions still remain about the nature of this system,
a periodicity of $\sim$ 5.5 yr in the X-ray, radio, mm, IR and optical
is now clearly established \citep[][and references
therein]{Damineli:2008b}, and has provided strong evidence for the
presence of a binary companion. With an orbital eccentricity of
$\simeq 0.9$ \citep{Smith:2004}, the stellar separation at periastron
may only be $\simeq 1.6$ au \citep[with a semi-major axis of 16.64 au,
][]{Hillier:2001}. The radius of the primary star is unclear but it
may be as high as 1 au \citep{Damineli:1996}. The companion star has
remained very elusive, but is thought to have a mass $\sim 30 \Msol$,
and to be either an O star or a WR star \citep*[][hereafter
PC02]{Pittard:2002}. In this paper the more massive star, the LBV,
will be referred to as the primary, and the smaller companion will be
referred to as the secondary. Typical orbital and stellar parameters
for the binary system are noted in Table~\ref{tab:testparameters}.

While the binary model is now commonly accepted, the orientation of
the system, the mass-loss rates of the stars, and the key physics
occuring around periastron passage are much more controversial. Most
works favour the secondary star moving behind the primary around
periastron \citep[e.g.][]{Damineli:1996, Pittard:1998, Corcoran:2001,
Corcoran:2005, Akashi:2006, Hamaguchi:2007, Nielsen:2007, Henley:2008,
Okazaki:2008}, though some others favour a system orientated such that
the secondary star is positioned in front of the primary during
periastron passage \citep{Falceta:2005, Kashi:2007}, or at quadrature
\citep{Ishibashi:2001, Smith:2004}. Estimates of the mass-loss rate of
the primary star range from a few times $10^{-4}$
\citep{Corcoran:2001, Pittard:2002} to a few times $10^{-3} \Msolpyr$
\citep{Hillier:2001, vanBoekel:2003}. Observations of the reflected
emission from the Homunculus by \cite{Smith:2003b} revealed a higher
degree of absorption at high latitudes, implying a higher rate of
mass-loss from the primary star near the poles. In contrast, the X-ray
emission from the WCR is most sensitive to the mass loss near the
orbital plane. If the orbital plane is aligned with the equatorial
region of the winds this may explain some of the difference in the
estimated mass-loss rates.

The X-ray emission has proved to be extremely useful for deciphering
the complex phenomena of $\eta\;$Car, because it is relatively
un-attenuated and can be used to probe deep into the centre of the
system. The keV X-ray emission displays a dominantly thermal spectrum,
and indicates the presence of hot shocked gas at a temperature of
$\sim 10^{7} - 10^{8} \rm{K}$. While there are several possible
methods for generating high temperature plasma, the X-ray lightcurve
provides strong support to the binary scenario, where the X-ray
emission naturally arises from the high-speed collision of the winds
in a massive binary system
\citep[e.g.][]{Stevens:1992,Pittard:1997,Pittard:2007}. Since the LBV
wind is slow, and cannot be shock heated to high temperatures, the
X-ray emission in this scenario must arise from a much faster
companion wind.

Fits to a grating spectrum from the \textit{Chandra} satellite
obtained mass loss rates and terminal wind speeds of $2.5\times10^{-4}
\Msolpyr$ and $500 \thinspace \rm{km\thinspace s}^{-1}$ for the
primary star, and $1.0\times10^{-5} \Msolpyr$ and $3000 \thinspace
\rm{km\thinspace s}^{-1}$ for the secondary star
\citep{Pittard:2002}. The spectrum was attained near apastron, when 2D
hydrodynamic simulations (which ignore orbital motion) give a good
description of the WCR.

A particularly interesting feature of the X-ray emission is the long
and deep minimum which coincides with periastron passage of the
stars. During this interval emission from the system undergoes
dramatic variations across a broad range of wavelengths, with high
excitation lines displaying a rapid reduction or ``collapse'' in
intensity \citep{Damineli:2008a}. We construct a 3D model of the WCR
in which rapid and large changes in the structure occur during
periastron passage. The model allows us to investigate the X-ray
emission throughout the entire orbit, so that the orbital orientation,
the eccentricity, and the wind parameters of the stars can be
determined. We then explore the physics of the WCR during periastron
passage. The layout of this paper is as follows:
\S~\ref{sec:min_intro} discusses a number of scenarios which have been
suggested to explain the behaviour of the X-ray minimum;
\S~\ref{sec:model} contains a brief description of the dynamic model
developed by \citet[][hereafter PP08]{Parkin:2008} which we apply to
\etacar in \S~\ref{sec:results}; \S~\ref{sec:minimum} examines a
number of mechanisms which may be important for bringing about the
X-ray minimum; and \S~\ref{sec:conclusions} concludes and outlines
possible future directions.

\vspace{-7mm}

\section{The X-ray minimum}
\label{sec:min_intro}

\subsection{Observations}

The \textit{Rossi X-ray Timing Explorer (RXTE)} has observed \etacar
since 1996, so that data over more than 2 orbital cycles is now
available. The X-ray lightcurve is shown in
Fig.~\ref{fig:rxtecurve}. Following apastron, there is a gradual
increase in emission over approximately 2 years, followed by a rapid
decline to a deep minimum which lasts for approximately 60 days
($\simeq 0.03 $ in orbital phase), then a shallower egress out of the
minimum to a roughly constant luminosity prior to its gradual rise
again in the next cycle \citep{Ishibashi:1999, Corcoran:2001,
Corcoran:2005}.

\textit{XMM-Newton} spectra analysed by \cite{Hamaguchi:2007} showed
that the flux in the 2-10 keV range at the start of the minimum was
0.7\% of the maximum value observed by \textit{RXTE} before the
minimum. Suprisingly, the flux in the latter half of the minimum
increases by a factor of 5 from the lowest observed value, indicating
that the minimum has two states. Some of the observed X-ray flux
during the minimum is emission from an earlier orbital phase which is
scattered by the Homunculus \citep{Corcoran:2004}.

\subsection{Possible explanations}
\label{subsubsec:possibleexplanations}

There are currently a number of possible explanations for the minimum
in the X-ray lightcurve which may, or may not, be mutually exclusive:
\begin{itemize}
\item \textit{The eclipse model}: The region of the WCR responsible
for emitting 0.1-10 keV X-rays may be fully occulted by the primary
star itself, or by its dense, optically thick wind (i.e. a wind
eclipse) if the secondary moves behind the primary star at
periastron. The lack of variation in the plasma temperature measured
in the \textit{XMM-Newton} spectra during the minimum is supportive of
an eclipse of the source, in which case residual emission arises from
regions further downstream. \cite{Okazaki:2008} find that pile-up of
the dense primary wind after periastron can cause an extended minimum
via a wind-eclipse. On the other hand, the deformation of the Fe XXV
profile and the relatively weak fluorescence Fe line intensity during
the minimum may suggest an intrinsic fading of the X-ray emissivity
\citep{Hamaguchi:2007}.
\item \textit{Increased mass-loss}: LBV's exhibit variability on a
number of timescales \citep[e.g.][]{deGroot:2001}, and are known to go
through eruptive phases where large amounts of stellar material are
ejected into the surrounding interstellar medium. The primary star may
be undergoing ${\rm S\;Dor}$-like fluctuations
\citep{vanGenderen:2007}, which may reach a maximum around periastron
with a shell-ejection event \citep[such an event was proposed by][as
an alternative to the binary eclipse model for explanation of the
X-ray minimum]{Davidson:1997}.

\begin{table*}
\begin{center}
\caption[]{Assumed parameters for $\eta\;$Car. $\eta =
\dot{M}_{2}v_{\infty 2}/ \dot{M}_{1}v_{\infty 1}$ is the wind momentum
ratio and $a$ is the semi-major axis of the orbit. References are as
follows: 1 = \cite{Davidson:1997}, 2 = \cite{Hillier:2001}, 3 =
\cite{Pittard:2002}, 4 = \cite{Corcoran:2005}, 5 =
\cite{Corcoran:2007}. After an extensive parameter space exploration
we found that a composite (ISM + nebula) column density (between the
observer at Earth and the edge of the simulation box,
$~1500\;{\rm au}$ from the stars) of $1\times10^{22}\;{\rm
cm}^{-2}$ provided better fits to the X-ray flux in the soft band (2-5
keV). This value is consistent with, but slightly lower than, the
value of $5\times10^{22}\;{\rm cm}^{-2}$ derived by
\cite{Hamaguchi:2007} for the ``the absorption beyond the central
source'', which likely represents the absorption column between the
secondary star at apastron and the observer at earth. $R_{\ast}$ is
taken to be the radius of the gravitationally bound region of the
star; for the primary star the photosphere will exist somewhere in the
stellar wind.}
\begin{tabular}{lllll}
\hline
Parameter & Primary & Secondary & System & Reference \\
\hline
$M$ (M$_{\odot}$) & 120 & 30 &  & 2 \\ 
$R_{\ast}$ (R$_{\odot}$) & 100 & 20 & & 5 \\
\hline
$\dot{M}$ ($10^{-5}\rm{M}_{\odot} \rm{yr}^{-1}$) & 25 & 1  & & 3\\
$v_{\infty}$ (km\thinspace s$^{-1}$) &  500 & 3000 & & 3 \\
$\eta$ & & & 0.24 & 3 \\
\hline
$a$ (au) & & & 16.64 & 2 \\
Orbital period (d) & & & 2024 & 4 \\
Eccentricity ($e$) & & & 0.9 & 4 \\
Distance (kpc) & & & 2.3 & 1 \\
ISM + nebula column ($10^{22}\rm{cm}^{-2}$) & & & 1 & \\
\hline
\label{tab:testparameters}
\end{tabular}
\end{center}
\end{table*}

\begin{figure}
\begin{center}
\psfig{figure=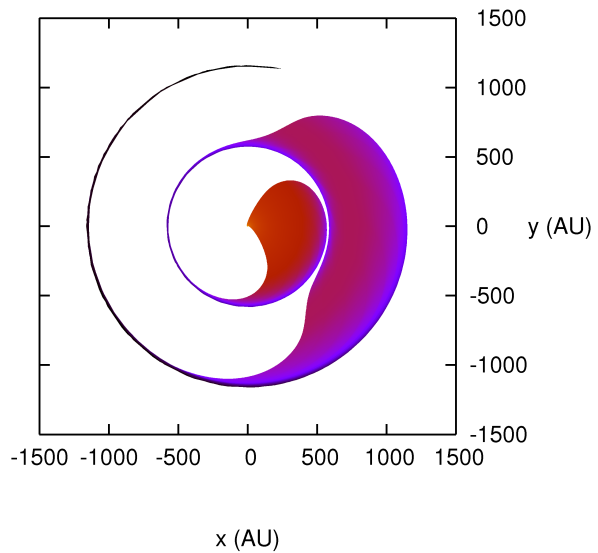,width=8.0cm}
\caption[]{The large scale structure of the contact discontinuity
(which separates the two winds) viewed from above the orbital
plane. The white region marks the unshocked wind of the primary
star. In this model the unshocked secondary wind is entirely contained
within the coloured region. The primary star is located at the origin
and the companion star orbits about this point. The WCR forms a spiral
structure due to the orbital motion of the stars. Asymmetry in the
distribution of the unshocked companion wind is caused by the highly
eccentric ($e = 0.9$) orbit. The thinning of the coloured region
towards the tail of the spiral is an artifact of the model (see PP08),
due to the fact that the WCR is traced back for only two orbits. The
parameters used in this simulation are noted in
Table~\ref{tab:testparameters}.}
\label{fig:spiral}
\end{center}
\end{figure}

\begin{figure*}
\begin{center}
    \begin{tabular}{llll}
      \resizebox{40mm}{!}{{\includegraphics{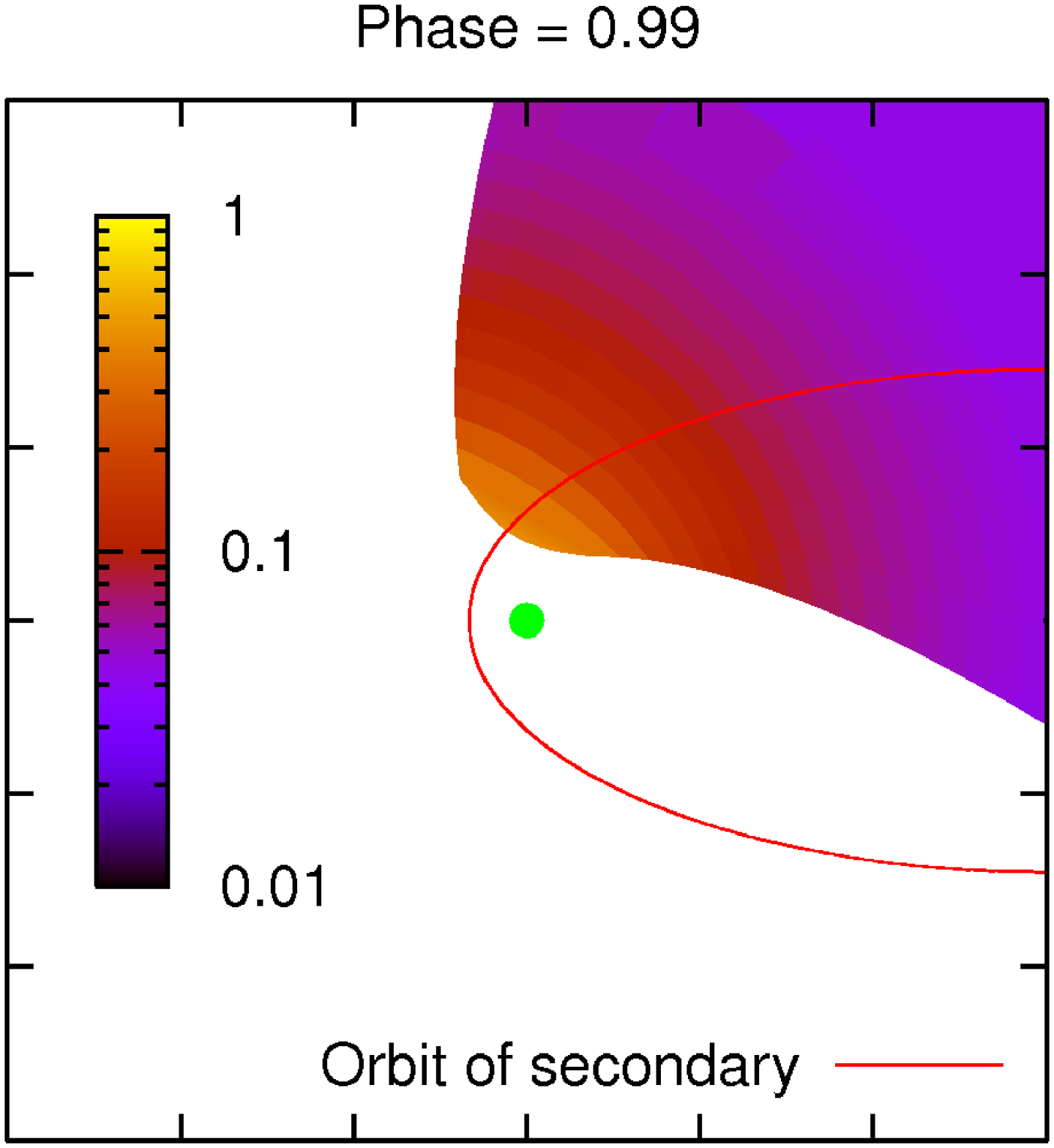}}} &
      \resizebox{40mm}{!}{{\includegraphics{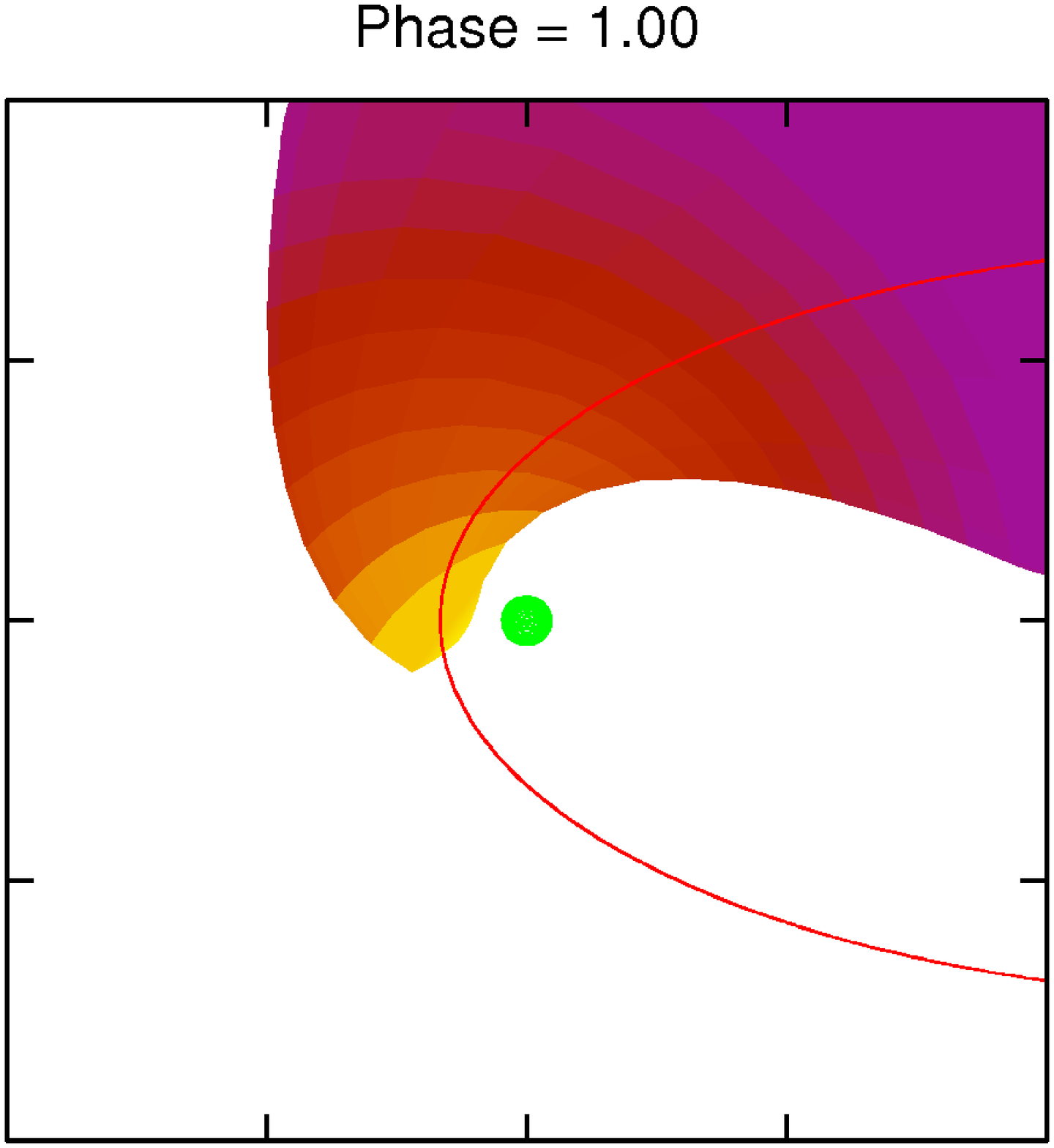}}} &
      \resizebox{40mm}{!}{{\includegraphics{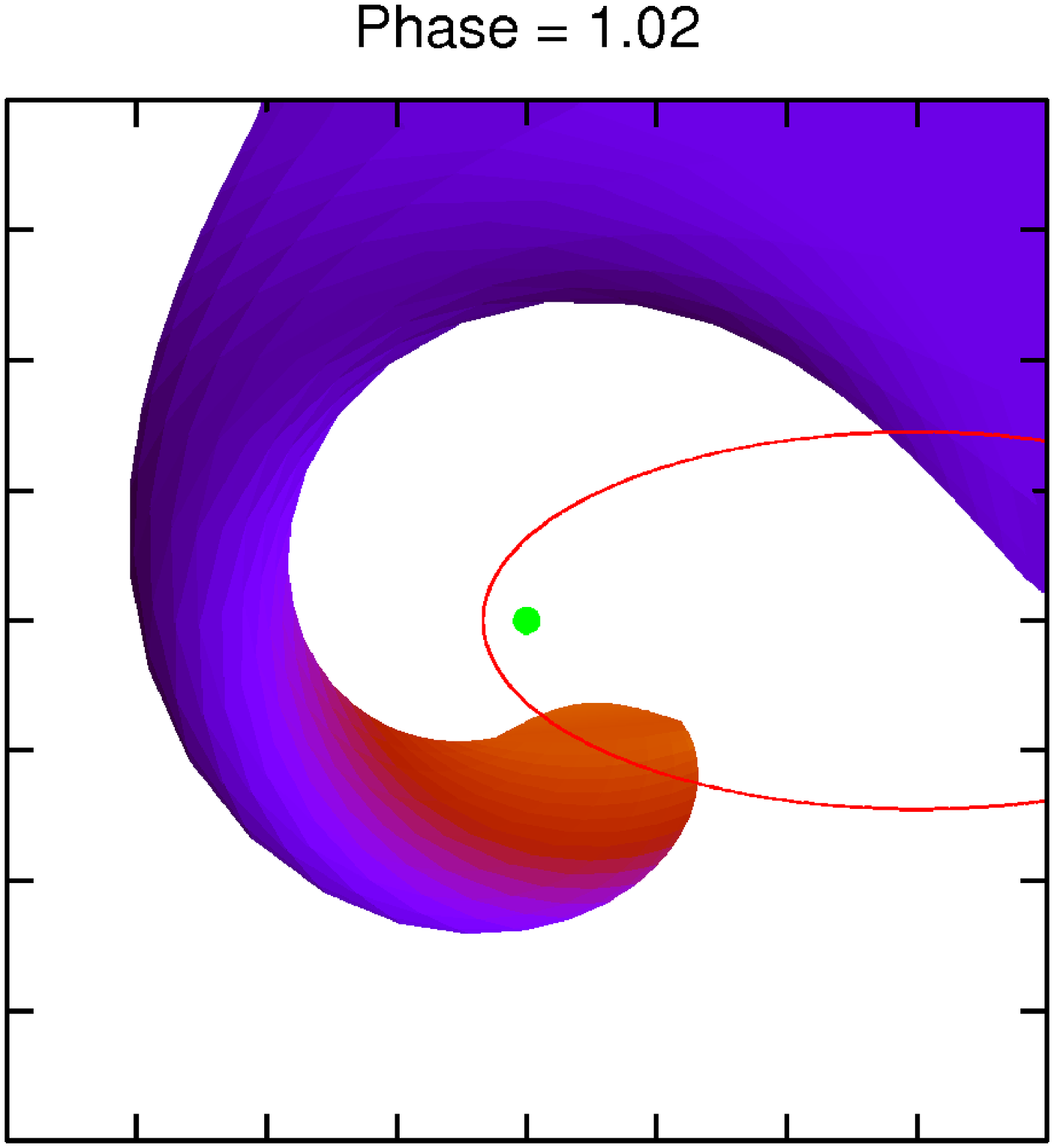}}} &
      \resizebox{40mm}{!}{{\includegraphics{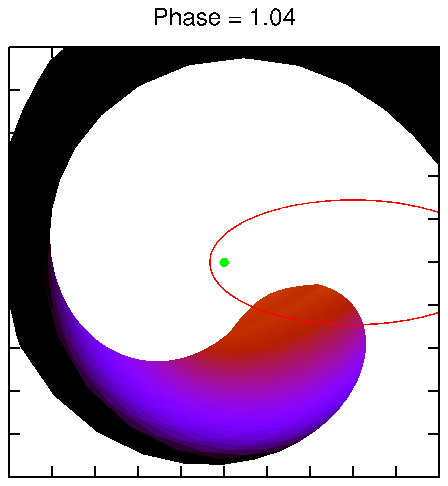}}} \\
    \end{tabular} 
\caption[]{The structure and position of the WCR around periastron
passage. The colour scale shows the surface density of the postshock
primary wind in g\thinspace cm$^{-2}$. The primary star is shown as a
green circle. The tick marks on the axis mark a distance of 5 au (note
the different scale of each panel). From left to right the plots
correspond to 20 days before periastron, periastron itself, and 40 and
81 days after periastron, and compare well against the SPH model of
\cite{Okazaki:2008}.}
\label{fig:clperi}
\end{center}
\end{figure*}

Theoretical work on tidal interactions in binary systems shows that
tidal oscillations may result in a phase of increased non-spherical
mass-loss \citep[][and references therein]{Moreno:2005,
Kashi:2008}. \cite{Corcoran:2001} found that incorporating a factor of
$\sim20$ increase in the primary star mass-loss rate for
$\delta\phi\sim0.04$ improved the fit to the duration of the
minimum. High velocity absorption lines ($\sim 750\;\rm{km\thinspace
s^{-1}}$ for He I $\lambda 6678$) observed by \cite{Stahl:2005} around
periastron in the reflected emission from the Homunculus nebula may be
evidence of a period of increased mass-loss.

\item \textit{Shut-down of the companion's wind}: If the WCR moves
into the acceleration region of the secondary's wind (due to
contraction of the stellar separation), a stable momentum balance may
be lost, resulting in the primary's wind overwhelming the secondary's
and causing the WCR to collapse onto the surface of the secondary
star. This may be aided by the radiative inhibition
\citep{Stevens:1994} or braking \citep*{Gayley:1997} of the secondary
star's wind at periastron due to the high stellar luminosity of the
primary star ($\simeq$ a few $\times 10^{6}\Lsol$). Evidence for a
collapse of the WCR onto the secondary star exists in the behaviour of
the highly ionised lines around the minimum
\citep{Damineli:2008a}. This possibility, and the potential occurance
of accretion of the primary wind by the companion star, was explored
by \cite{Akashi:2006}.
\end{itemize}

\section{The dynamic model}
\label{sec:model}

To successfully model the X-ray minimum the motion of the shocked and
unshocked gas relative to the stars must be taken into
account. Unfortunately, with current computational resources, it is
difficult to explore a wide range of parameter space with 3D models,
even when using sophisticated grid or particle based hydrodynamic
codes. The high orbital eccentricity of \etacar compounds this
problem, and 3D simulations of \etacar have only examined a small
region of parameter space to date \citep{Pittard:2000thesis,
Okazaki:2008}. While both of these efforts revealed important insights
into the dynamics of the WCR, only a small part of the orbit around
periastron passage was simulated in \cite{Pittard:2000thesis}, while
the assumption of isothermality in \cite{Okazaki:2008} does not allow
direct calculation of the X-ray emission.

In this paper we use a recently developed 3D dynamical model which
allows efficient exploration of parameter space. A detailed
description can be found in PP08 where a demonstration with lower
eccentricity O+O and WR+O-star binaries is given. The model uses
equations from \cite{Canto:1996} to determine the ram pressure balance
and position of the contact discontinuity (CD) between the winds
(modified by an aberration angle due to orbital motion). The flow
downstream of the ``apex'' of the WCR is then assumed to travel
ballistically. The motion of the stars causes the contact surface to
wind up and form large scale spiral structures
(Fig.~\ref{fig:spiral}), reminiscent of the beautiful ``pinwheel
nebulae'' \citep{Tuthill:2008}. The shape of the WCR around periastron
passage is highly distorted due to the rapid transit of the stars and
the slow speed of the primary wind (Fig.~\ref{fig:clperi}).

The X-ray emission from the WCR is a function of the gas temperature
and density. Since the dynamical model does not contain such
information we use a grid-based, 2D hydrodynamical calculation of an
axis-symmetric WCR to obtain this. The resulting emission as a
function of off-axis distance is then mapped on to the coordinate
positions in the 3D dynamical model. In this way we obtain the benefit
of effectively modelling the thermodynamic and hydrodynamic behaviour
responsible for the production of the X-ray emission, while
simultaneously accounting for the effect of the motion of the stars on
the large-scale structure of the WCR and the subsequent wind
attenuation. Since the hydrodynamic calculation is 2D, the
computational requirements remain low.

The X-ray luminosity per unit volume is given by $\Gamma (E) =
n^{2}\Lambda(E,T)$, where $n$ is the gas number density ($\cm^{-3}$)
and $\Lambda(E,T)$ is the emissivity ($\rm{erg\; cm}^{3}s^{-1}$) as a
function of energy $E$ and temperature $T$ for optically thin gas in
collisional ionization equilibrium. $\Lambda(E,T)$ is obtained from
look-up tables calculated from the \textit{MEKAL} plasma code
\citep[][and references therein]{Liedahl:1995}, containing 200
logarithmically spaced energy bins in the range 0.1-10.0 keV, and 101
logarithmically spaced temperatures from $10^{4}$ to $10^{9}
\rm{K}$. Solar abundances are assumed.

The orbit of the stars is modelled in the $xy$ plane, so that viewing
angles into the system can be described by the inclination angle that
the line-of-sight makes with the $z$ axis, $i$, and the angle the
projected line-of-sight makes with the major axis of the orbit,
$\theta$ (positive values correspond to an angle subtended against the
positive $x$ axis in the prograde direction). The components of the
unit vector along the line-of-sight, $\underline{\hat{u}}$, are
\begin{eqnarray*}
   u_{x} & = & \cos\theta\sin i, \\
   u_{y} & = & \sin\theta\sin i, \\
   u_{z} & = & \cos i.
\label{eqn:losunitvectors}
\end{eqnarray*}

\noindent To determine the observed attenuated emission ray-tracing is
performed through the spiral structure. Specific details of the
emission and absorption calculations are now discussed.

\begin{figure*}
\begin{center}
    \begin{tabular}{ll}
      \resizebox{80mm}{!}{{\includegraphics{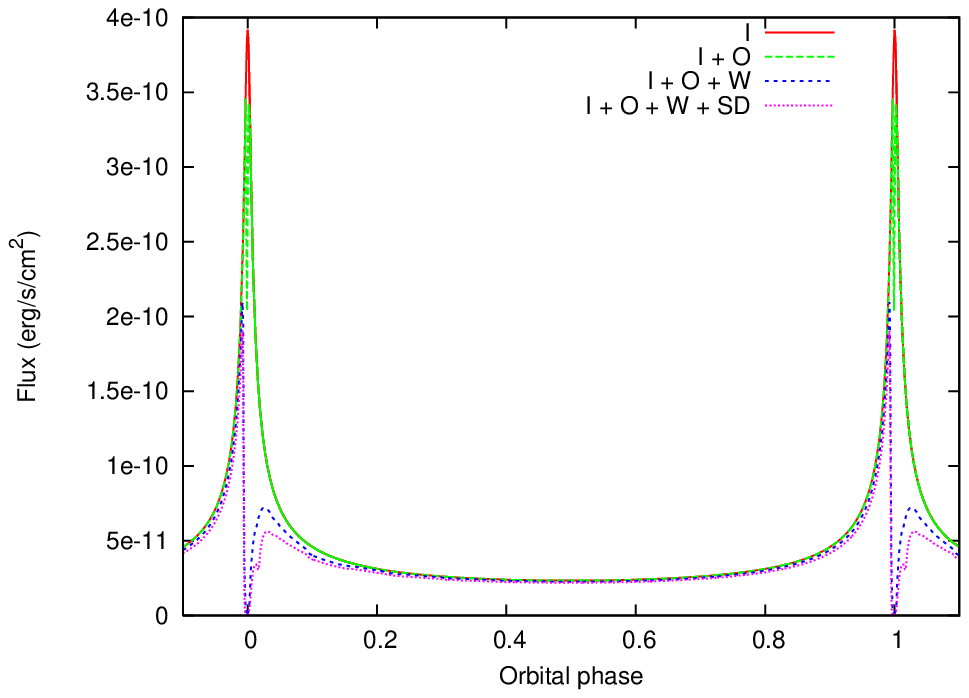}}} &
\resizebox{80mm}{!}{{\includegraphics{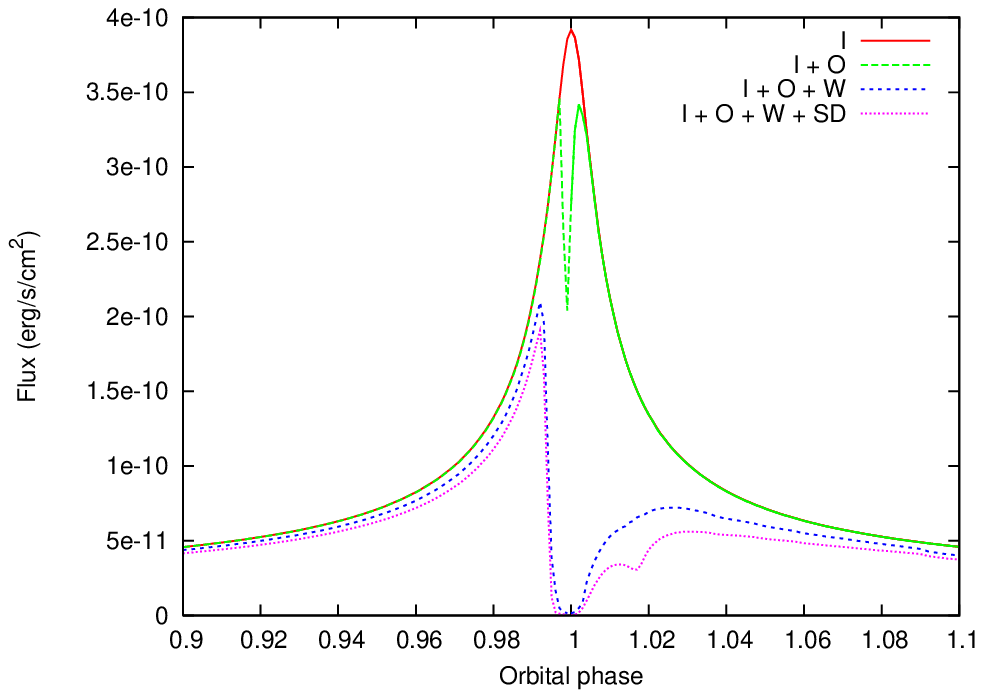}}} \\
    \end{tabular} 
\caption[]{The effect of various attenuation mechanisms on the
observed 2-10 keV emission as a function of orbital phase over an
entire orbital period (left), and over the orbital phase range
0.90-1.10 (right). The un-attenuated intrinsic emission (I) is shown
by the solid red curve. Various attenuation mechanisms are then
sequentially added: occultation (O), wind attenuation (W), and
attenuation through the thin dense postshock primary wind layer
(SD). The viewing angles for these curves are $i = 90^{\circ}$ and
$\theta = 0^{\circ}$, with the other model parameters listed in
Table~\ref{tab:testparameters}. The aberration of the WCR due to
orbital motion and interstellar + nebula absorption
($1\times10^{22}\rm{cm}^{-2}$) are included.}
\label{fig:abs_lc}
\end{center}
\end{figure*}

\subsection{Emission from the shocked secondary wind}
\label{subsec:emission}

The importance of cooling in the WCR can be quantified using the
cooling parameter \citep{Stevens:1992}

\begin{equation}
   \chi = \frac{t_{\rm{cool}}}{t_{\rm{esc}}} = \frac{v^4 _8
   d_{12}}{\dot{M} _{-7}},
\end{equation}

\noindent where $v_8$ is the preshock wind velocity in units of $1000
\thinspace\rm{km\thinspace s}^{-1}$, $d_{12}$ is the distance from the
star to the contact discontinuity in units of $10^{12}\rm{cm}$,
$\dot{M}_{-7}$ is the mass-loss rate of the star in units of $10^{-7}
\Msolpyr$, $t_{\rm{cool}}$ is the cooling time, and $t_{\rm{esc}}$ is
the escape time of gas out of the system ($\simeq d_{12} /
c_{\rm{s}},$ where $c_{\rm{s}}$ is the postshock sound
speed). $\chi_{1}$ and $\chi_{2}$ are the cooling parameter for the
postshock primary and secondary winds respectively.

In the case of $\eta\;$Car, the postshock primary wind is strongly
radiative ($\chi\simeq 10^{-4}-10^{-2}$), rapidly cools, and collapses
to form a thin dense sheet. As previously noted, this wind is not
expected to produce thermal X-rays above 1-2 keV as the relatively low
wind speed ($v_{1}\sim 500\thinspace\rm{km\thinspace s}^{-1}$) does
not shock-heat gas to high enough temperatures. In contrast, the high
velocity of the secondary's wind ($v_{2} \sim 3000 \thinspace
\rm{km\thinspace s}^{-1}$) causes its postshock flow to be adiabatic
($\chi\gtsimm 10$) around most of the orbit.  To first order the
entirety of the observable X-ray emission originates from the
postshock secondary wind although mixing at the CD due to dynamical
instabilities (PC02; see also Fig.~\ref{fig:hydro_cooling}) is a
complicating factor. Note that lower values of $\chi_{2}$ at
periastron indicate that cooling may be important for a short period
(Fig.~\ref{fig:cooling}), especially if the secondary wind is clumpy
and/or if radiative inhibition reduces its preshock speed. These
possibilities are investigated in more detail in
\S~\ref{sec:minimum}. In the meantime, we assume that the shocked
secondary's wind is adiabatic throughout the entire orbit, emits the
entirety of the X-ray emission, and that the intrinsic X-ray
luminosity scales as $1/d_{\rm{sep}}$, where $d_{\rm{sep}}$ is the
stellar separation \citep{Stevens:1992,Pittard:1997}.

\subsection{Attenuation}
\label{subsec:attenuation}

The intrinsic X-ray emission is attenuated by the unshocked winds. In
wide binaries one can usually assume that the winds are
instantaneously accelerated. However, in \etacar the primary's wind
may accelerate very slowly and this may affect the degree of
attenuation observed. When describing the wind acceleration using a
$\beta$-velocity law, a typical value for the slow acceleration of an
LBV wind would be $\beta = 4$ \citep{Barlow:1977, Pauldrach:1990}. To
test whether this is an important factor we compared simulations
adopting accelerating and instantaneously accelerated winds and found
that the difference in attenuation was small at all phases, and for
all lines of sight.

Attenuation by the shocked winds also occurs. Since the shocked
secondary wind is assumed to be hot, it is also assumed to have
negligible attenuation\footnote{This is a suitable assumption for the
purposes of our model, however it has been suggested by
\cite{Henley:2008} that attenuation by the companion wind plays a role
in shaping the emission line profiles.}. In contrast, the dense, cool
layer of postshock primary wind is a significant source of
absorption. The surface density of this layer changes by over two
orders of magnitude between periastron and apastron ($\sigma \propto
d_{\rm{sep}}^{-2}$), resulting in a dramatic increase in attenuation
around periastron for X-rays. For viewing angles closely tangential
with the shock surface the combination of the increased postshock gas
density and the pathlength dependent absorption described in PP08 is
sufficient to fully absorb X-rays with energies $\ltsimm 4\;$ keV.

\subsection{Observed emission}
\label{subsec:observed_emission}

\begin{figure}
\begin{center}
\psfig{figure=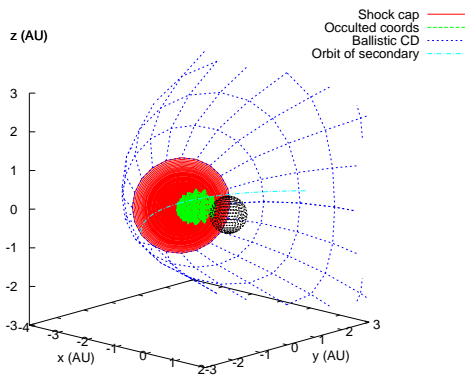,width=9.0cm}
\caption[]{A 3D plot of the occultation of the WCR by the primary star
at periastron ($\phi=0.0$) with $e = 0.9$ and a viewing angle of $i =
90^{\circ}$ and $\theta=0^{\circ}$. The region occulted by the primary
star (displayed in black) is shown in green. Although the primary
star is directly in front of the secondary star for the chosen values
of $i$ and $\theta$, due to aberration the shadow does not fall
directly on the apex of the WCR.}
\label{fig:occ_shkcaps}
\end{center}
\end{figure}

\begin{figure*}
\begin{center}
    \begin{tabular}{ll}
      \resizebox{80mm}{!}{{\includegraphics{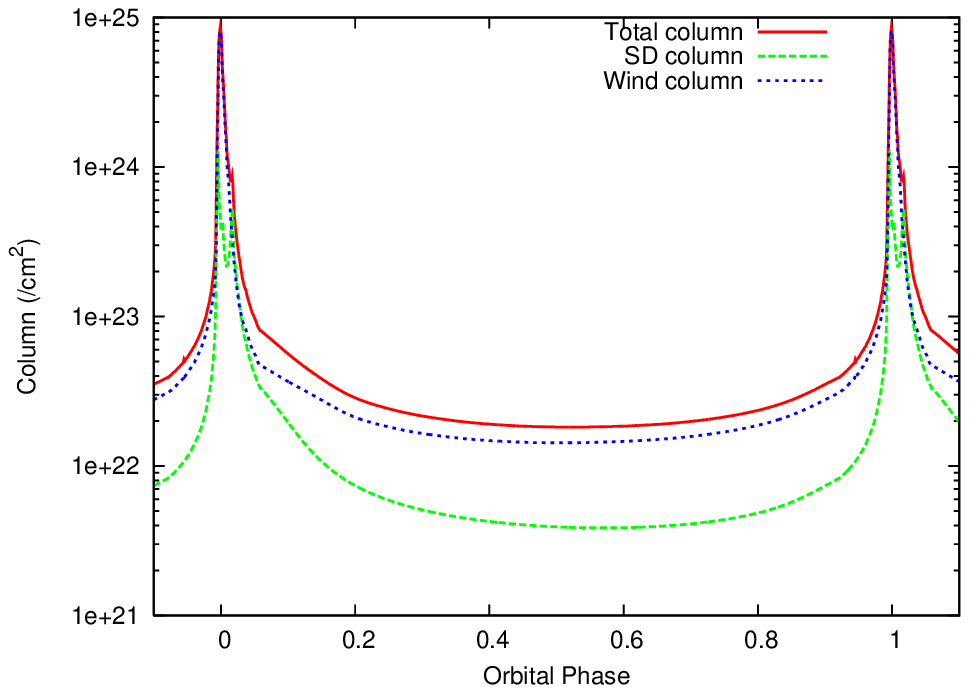}}} &
\resizebox{80mm}{!}{{\includegraphics{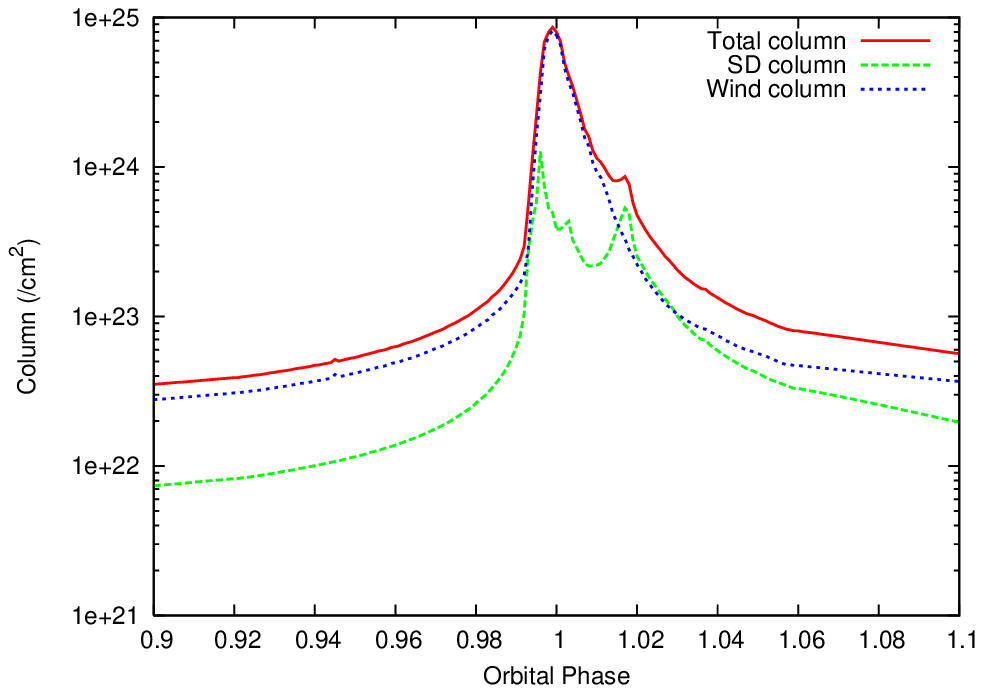}}} \\
    \end{tabular} 
\caption[]{The effect of various attenuation mechanisms on the
emission weighted column density as a function of orbital phase. The
total column (red), attenuation due to the unshocked stellar winds
(blue), and attenuation due to X-rays intersecting the dense, cold
layer of postshock primary wind (green) are shown over an entire
orbital period (left) and over the orbital phase range 0.90-1.10
(right). Interstellar + nebula absorption ($\sim
1\times10^{22}\cm^{-2}$) has not been added to these plots. The
viewing angles are $i=90^{\circ}$ and $\theta = 0^{\circ}$.}
\label{fig:abs_col}
\end{center}
\end{figure*}

Prior to modelling the \textit{RXTE} lightcurve, we explore the effect
on the observed X-ray lightcurve of including various occultation and
attenuation mechanisms. A lightcurve from a model with the parameters
listed in Table~\ref{tab:testparameters} and viewing angles $i =
90^{\circ}, \theta = 0^{\circ}$ is shown in Fig.~\ref{fig:abs_lc}. The
skew of the WCR due to orbital motion causes the minimum to be
asymmetric, as observed. A particularly interesting feature is the
shoulder seen on the egress out of minimum. This is caused by
attenuation through the dense layer of shocked primary wind (see
below).

\begin{figure*}
\begin{center}
    \begin{tabular}{ll}
      \resizebox{80mm}{!}{{\includegraphics{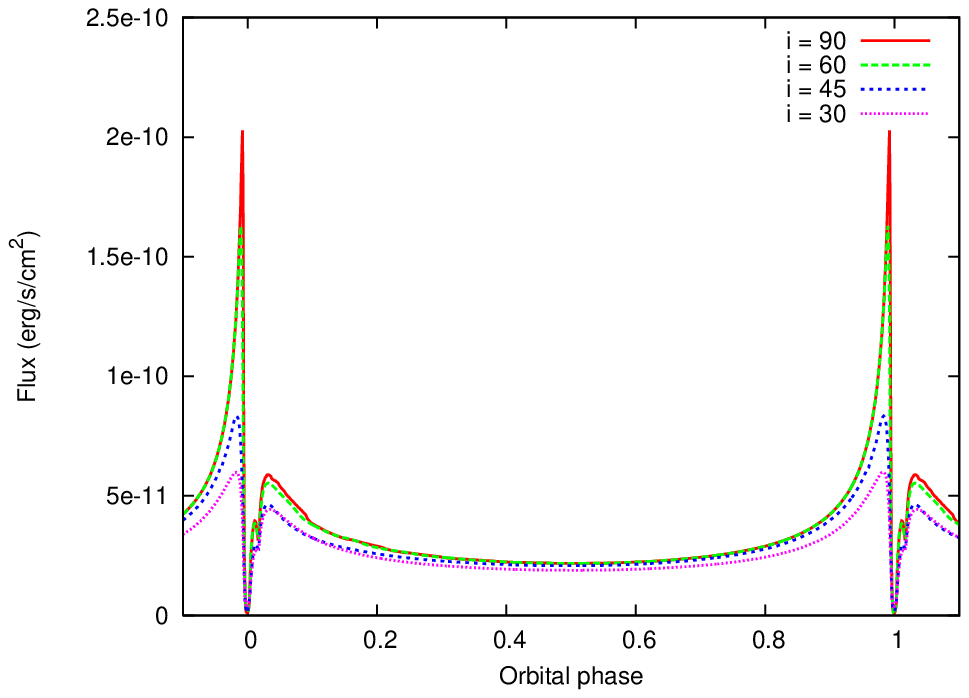}}} &
\resizebox{80mm}{!}{{\includegraphics{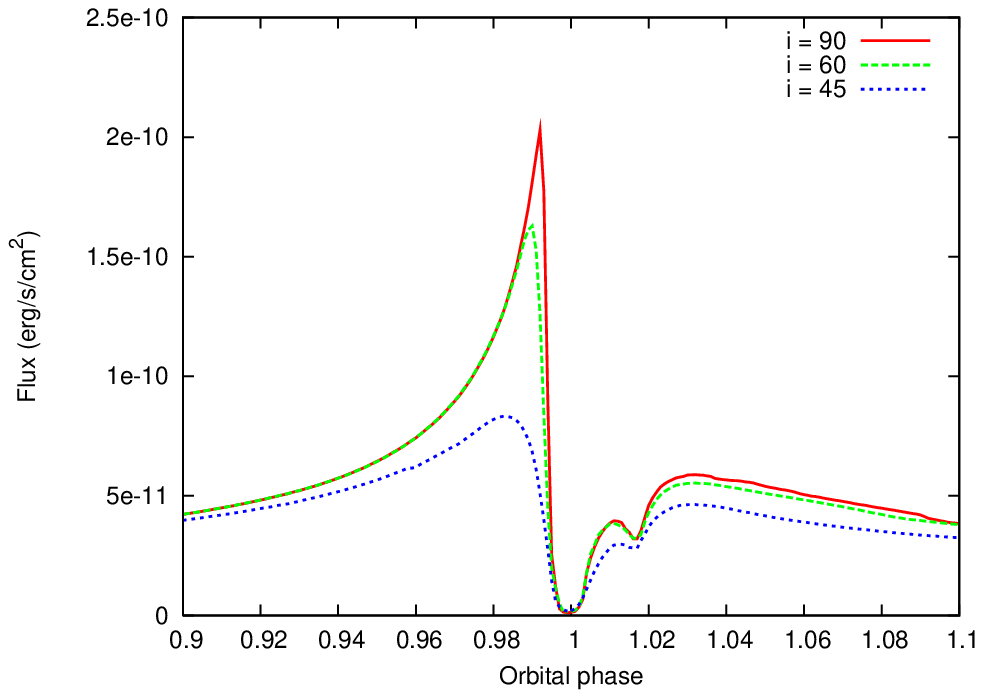}}} \\
    \end{tabular} 
\caption[]{Synthetic 2-10 keV lightcurves for various inclination
angles, $ i $, over an entire orbital period (left) and over the
orbital phase range 0.90-1.10 (right). An inclination angle of
$45^{\circ}$ provides the best resemblence to the observed morphology
(Fig.~\ref{fig:rxtecurve}). The parameters used to calculate these
results are noted in Table~\ref{tab:testparameters}. $\theta =
0^{\circ}$ in each of these models.}
\label{fig:inc_lc}
\end{center}
\end{figure*}

\begin{figure*}
\begin{center}
    \begin{tabular}{ll}
      \resizebox{80mm}{!}{{\includegraphics{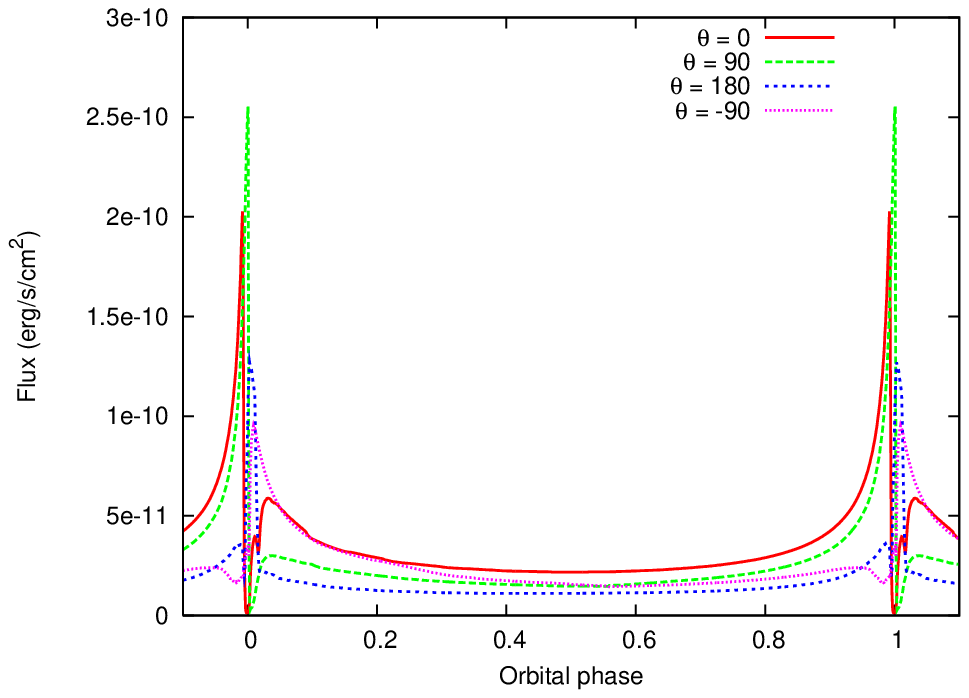}}} &
\resizebox{80mm}{!}{{\includegraphics{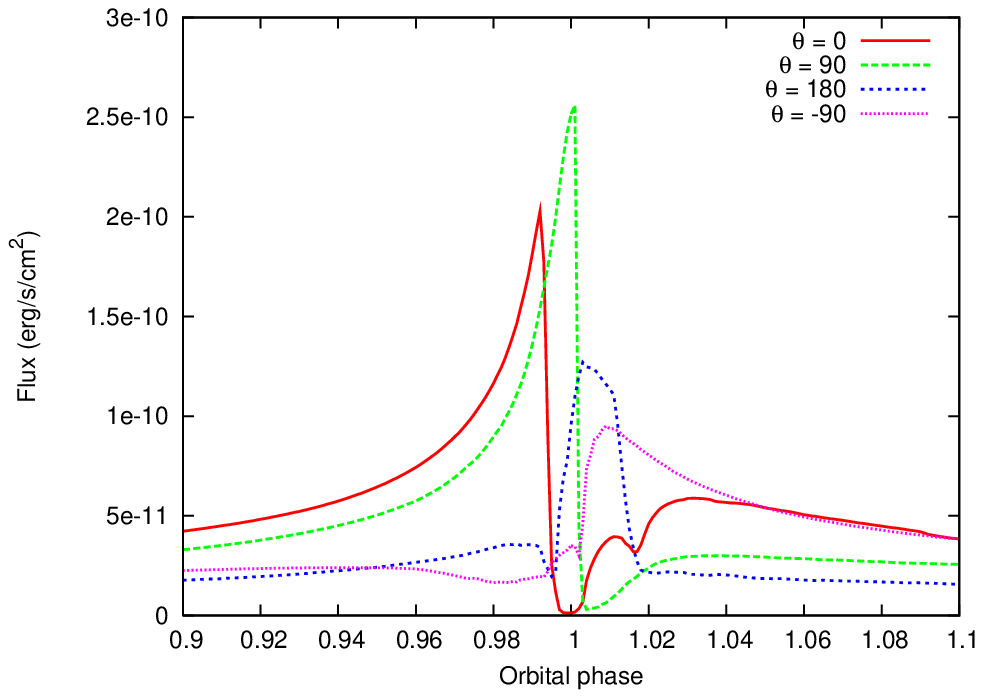}}} \\
    \end{tabular} 
\caption[]{Synthetic 2-10 keV lightcurves for an inclination $i =
90^{\circ}$, eccentricity, $e = 0.9$, and various line-of-sight
angles, $\theta$, over an entire orbital period (left) and over the
orbital phase range 0.90-1.10 (right). All other parameters are noted
in Table~\ref{tab:testparameters}.}
\label{fig:los_lc}
\end{center}
\end{figure*}

Occultation by the primary star (Fig.~\ref{fig:occ_shkcaps}) causes a
considerable reduction in emission around periastron though the effect
is very short-lived and does not cause the observed minimum (see
Fig.~\ref{fig:abs_lc}). Instead, the largest source of attenuation
arises from the unshocked stellar winds except for a phase interval of
$\approx 0.01$ during the minimum when absorption through the shocked
primary wind dominates (Fig.~\ref{fig:abs_col}). The smallest values
of the emission weighted column (PP08) occur at apastron, when the
sight lines from the head of the WCR initially pass through the
secondary wind. A circumstellar column of $2\times10^{22}\rm{cm}^{-2}$
(the absorption from shocked and unshocked winds within the simulation
box, see PP08), plus a column of $1\times10^{22}\rm{cm}^{-2}$ which
includes absorption in the Homunculus and ISM, is sufficient to absorb
$\sim 99.99\%, 90\%,$ and $20\%$ of the 1, 2, and 5 keV X-rays from
the WCR respectively. The column density begins to increase as the
stars approach periastron, then there is a dramatic rise in the column
through the unshocked winds at $\phi \sim 0.99$ which highlights the
shift of the lines-of-sight into the high density primary wind. This
is accompanied by a peak in absorption by the shocked primary wind,
represented by the surface density (SD) component. The two peaks in
the SD component column at phases $\phi \sim 0.99$ and 1.02 are
produced by sight lines originating from points with the highest
intrinsic luminosity becoming closely tangential with the shock
surface. The short duration in phase of the peak in column density
(which briefly reaches $\sim 10^{25}\rm{cm}^{-2}$) denotes the rapid
motion of the stars through periastron passage.

The attenuation along lines-of-sight close to the positive x-axis
(i.e. $i=90^{\circ}$, $\theta=0^{\circ}$) at phases near apastron is
composed of large volumes of unshocked secondary wind and slices of
dense primary wind (c.f. Fig.~\ref{fig:spiral}), with the former
dominating.

Increasing the radius of the primary star causes the depth, and width,
of the occultation component of the minimum, and the actual minimum to
increase slightly, but the high eccentricity combined with the rapid
orbital motion of the stars around periastron results in very little
change to the attenuating column densities. The lightcurve is also
relatively insensitive to the rate of the acceleration of the primary
wind (which affects the density near to the star) and to the velocity
cut at which the flow in the WCR is assumed to behave ballistically
(see PP08).

\section{Results}
\label{sec:results}

\subsection{Varying model parameters}
\label{subsec:modelparameters}

In the following subsections we explore the effect of varying the
orientation of the orbit, the wind momentum ratio, and the
eccentricity. 

\subsubsection{Variation with orbital inclination}

Fig.~\ref{fig:inc_lc} shows how varying the orbital inclination angle
changes the observed lightcurve. For values of $i = 60^{\circ} -
90^{\circ}$ the synthetic lightcurves are largely similar. For $i
\simeq 45^{\circ}$, similar to the $42^{\circ}$ inclination angle
derived for the polar axis of the Homunculus nebula
\citep{Smith:2006}, the ratio of the pre- and post-minimum X-ray
emission is reduced considerably and is in much better agreement with
the \textit{RXTE} data \citep[see also ][]{Steiner:2004, Akashi:2006,
Kashi:2007, Okazaki:2008}. Reducing the inclination angle further to
$i = 30^{\circ}$ causes the pre-/post-minimum flux ratio to become too
small. Although reducing $i$ decreases the direct occultation of the
region of the WCR with the largest intrinsic luminosity around
periastron passage (Fig.~\ref{fig:occ_shkcaps}), the wind absorption
increases as a larger fraction of the X-rays initially pass through
the primary wind (cf. Fig.~\ref{fig:inc_lc}).

\begin{table}
\begin{center}
  \caption[]{Parameters used for the models shown in
Figs.~\ref{fig:results_lc_eta} and \ref{fig:results_col}. In all
simulations the terminal velocity of the primary, $v_{\infty 1}$, and
secondary, $v_{\infty 2} $, stars wind's are 500 and
$3000\;\rm{km\thinspace s}^{-1}$ respectively.}
\begin{tabular}{lll}
\hline
 $\eta$ & $\dot{M_{1}}$ & $\dot{M_{2}}$ \\
 & ($M_{\odot} \rm{yr}^{-1}$) & ($M_{\odot} \rm{yr}^{-1}$)  \\
\hline
0.24 &  $3.5\times10^{-4}$  & $1.4\times10^{-5}$  \\
0.18 &  $4.7\times10^{-4}$  & $1.4\times10^{-5}$  \\
0.12 &  $7.0\times10^{-4}$  & $1.4\times10^{-5}$  \\
0.08 &  $1.1\times10^{-3}$  & $1.4\times10^{-5}$  \\
0.04 &  $2.1\times10^{-3}$  & $1.4\times10^{-5}$  \\
0.024 & $3.5\times10^{-3}$ & $1.4\times10^{-5}$  \\
\hline
\label{tab:results}
\end{tabular}
\end{center}
\end{table}

\begin{figure}
\begin{center}
    \begin{tabular}{l}
      \resizebox{80mm}{!}{{\includegraphics{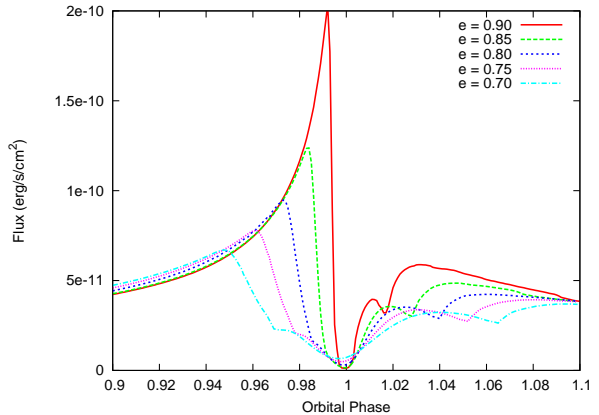}}} \\
    \end{tabular} 
\caption[]{Synthetic 2-10 keV lightcurves over the orbital phase range
  0.90-1.10 for $\eta = 0.24$, an inclination $i = 90^{\circ}$,
  line-of-sight angle $\theta = 0^{\circ}$, and varying values for
  $e$. The orbital and wind parameters used are noted in
  Table~\ref{tab:testparameters}.}
\label{fig:ecc_lc}
\end{center}
\end{figure}

\begin{figure}
\begin{center}
    \begin{tabular}{l}
      \resizebox{80mm}{!}{{\includegraphics{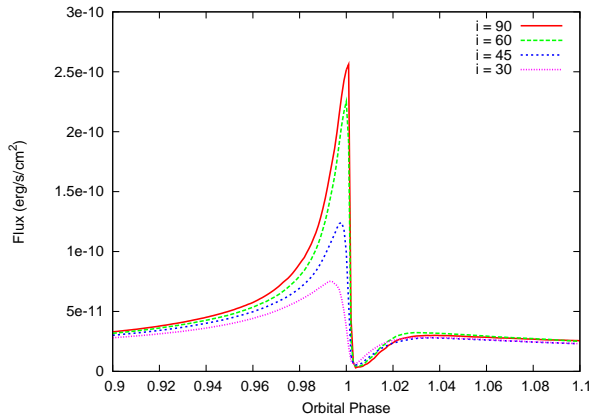}}} \\
    \end{tabular} 
\caption[]{Synthetic 2-10 keV lightcurves for $\eta = 0.24$,
  line-of-sight angle $\theta = 90^{\circ}$, eccentricity $e = 0.9$,
  and varying inclination angle, $i$, over the orbital phase range
  0.90-1.10. The orbital and wind parameters used to calculate these
  results are noted in Table~\ref{tab:testparameters}. Comparison with
  Fig.~\ref{fig:rxtecurve} shows that these curves do not reproduce
  the observed lightcurve.}
\label{fig:los_lc_theta90}
\end{center}
\end{figure}

\subsubsection{Variation with the orientation of the orbital semi-major axis}
\label{subsubsec:los}

As mentioned in \S~\ref{sec:intro}, there are competing suggestions
for the orientation of the semi-major axis. Fig.~\ref{fig:los_lc}
shows the effect on the observed X-ray emission of varying the
line-of-sight angle, $\theta$, while keeping the inclination angle
constant at $i = 90^{\circ}$. In each case, the maximum emission
occurs while the sight line to the apex of the WCR is through the less
dense secondary wind. 

\subsubsection{Variation with orbital eccentricity}

For the majority of the calculations an orbital eccentricity of $e =
0.9$ was adopted. The eccentricity affects the orbital velocity of the
companion star around periastron, the distance of closest approach of
the stars, and the amount of time the companion star and the region of
the WCR with the highest intrinsic luminosity (assuming $L_{X_{\rm
    int}} \approx 1/d$, but see \S~\ref{sec:minimum} for caveats)
spends immersed in the dense primary wind. Lowering the value of $e$
increases the width of the minimum towards $\phi \sim 0.03$ which is
desired (Fig.~\ref{fig:ecc_lc}). However, the decline to minimum then
occurs at earlier phases. This can be countered by increasing the
value of $\theta$ (c.f. Fig.~\ref{fig:los_lc}), although there are
then implications for the pre-/post-minimum luminosity ratio, and the
morphology of the egress out of minimum. On the other hand, increasing
the eccentricity decreases the duration of the minimum and increases
the pre-/post-minimum luminosity ratio. Overall, a best-fit is
obtained with $e \simeq 0.9$, in agreement with previous work
\citep[e.g.][]{Henley:2008}. However, if the WCR collapses at
periastron (see \S~\ref{sec:minimum}) the eccentricity may be higher.

\subsubsection{Variation with wind momentum ratio}
\label{subsubsec:varying_eta}

\begin{figure*}
\begin{center}
    \begin{tabular}{ll}
      \resizebox{80mm}{!}{{\includegraphics{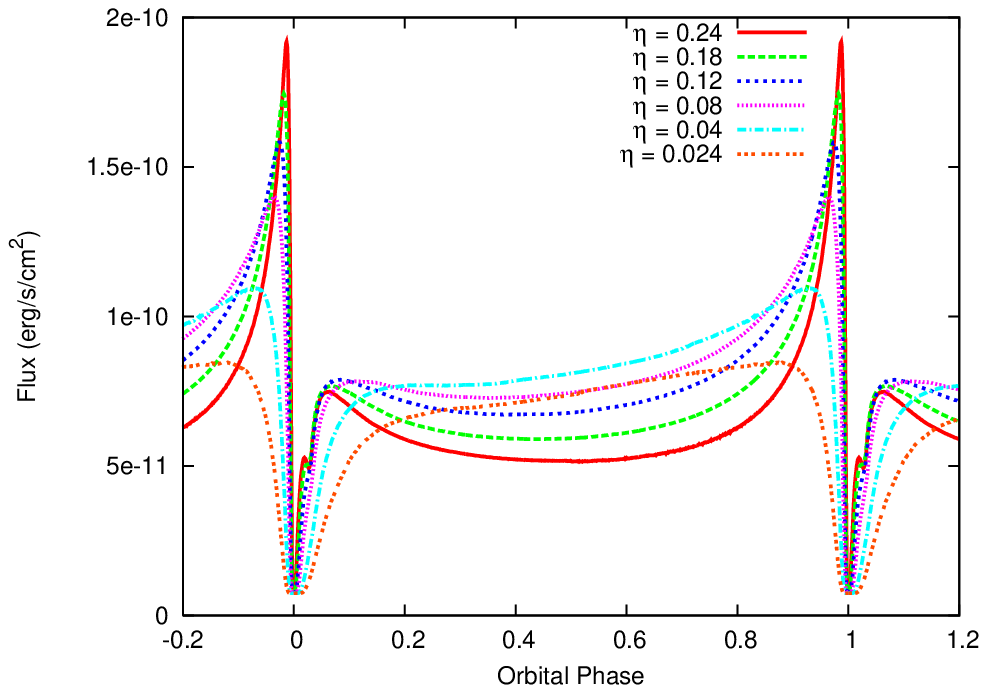}}} &
\resizebox{80mm}{!}{{\includegraphics{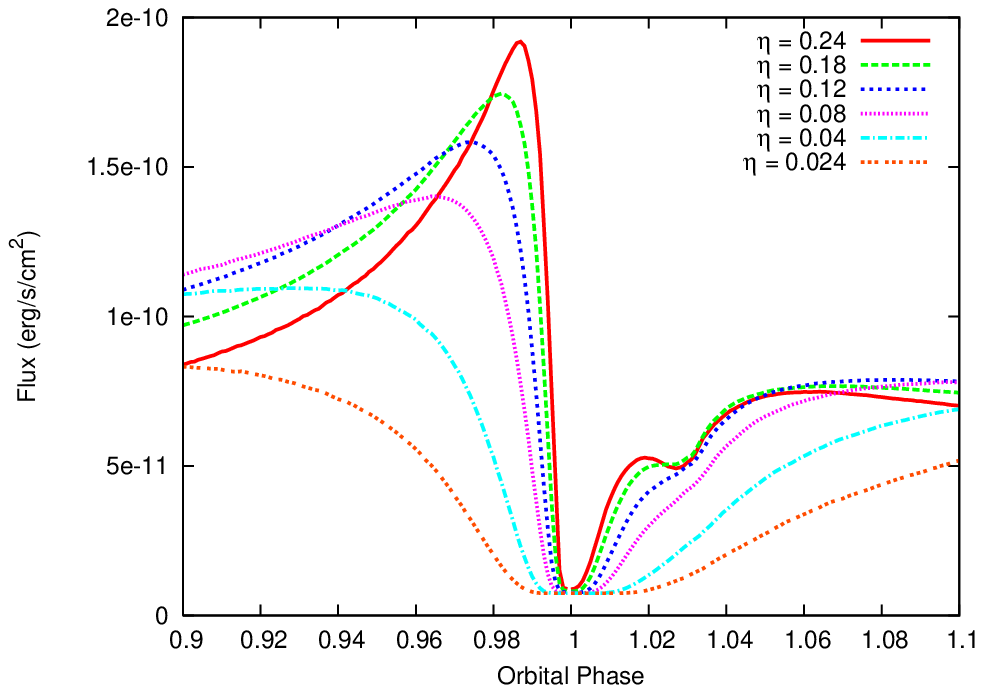}}} \\
    \end{tabular} 
\caption[]{Variation of the attenuated 2-10 keV X-ray emission for $e
  = 0.9$, $ i = 42^{\circ}$, $\theta = 20^{\circ}$ and varying values
  of the wind momentum ratio $\eta $ shown over the whole orbit (left)
  and over the orbital phase range 0.90-1.10 (right). The orbital and
  wind parameters are noted in Tables~\ref{tab:testparameters} and
  \ref{tab:results}.}
\label{fig:results_lc_eta}
\end{center}
\end{figure*}

Varying the momentum ratio between the winds (by adjusting
$\dot{M}_{1}$) affects the opening angle of the WCR and the spatial
distribution of the unshocked winds, and changes the attenuation and
observed X-ray emission, as shown in
Fig.~\ref{fig:results_lc_eta}. Decreasing $\eta$ causes the sight
lines to the apex of the WCR to move into the denser LBV wind at
earlier orbital phases, and produces a wider minimum at
periastron. When $\eta=0.024$ the opening angle of the WCR is reduced
so much that the sight line from the WCR apex moves into the primary
wind at such an early phase that a significant rise in X-ray
luminosity before periastron is prevented. Another effect of reducing
$\eta$ is that more of the secondary wind becomes shocked in the
WCR. This means that the luminosity at apastron initially increases as
$\eta$ decreases from 0.24. However, the opening angle of the WCR soon
decreases to such an extent that sightlines to the apex of the WCR
move into the primary wind, so that the luminosity at $\phi=0.5$
declines for $\eta \ltsimm 0.04$.

\begin{figure*}
\begin{center}
    \begin{tabular}{ll}
      \resizebox{80mm}{!}{{\includegraphics{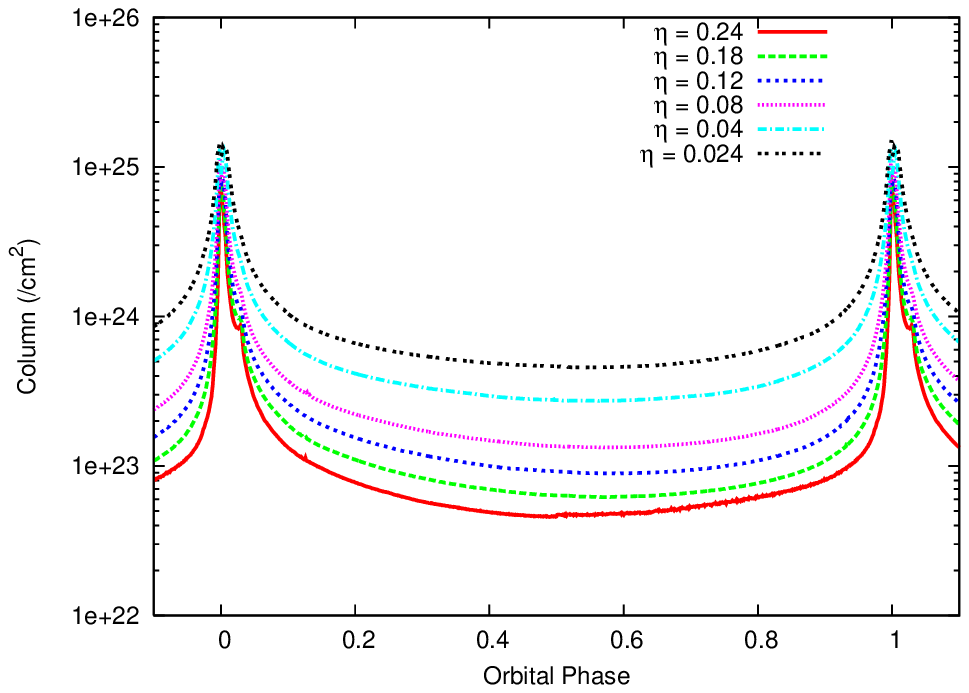}}} &
\resizebox{80mm}{!}{{\includegraphics{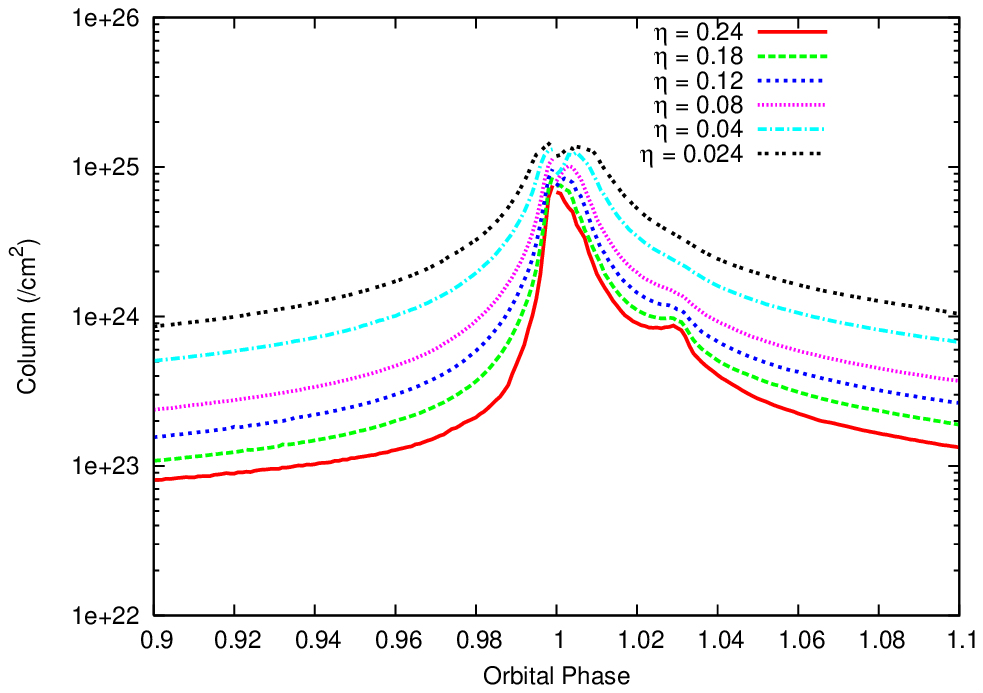}}} \\
    \end{tabular} 
\caption[]{Variation of the emission weighted column density
($\cm^{-2}$) for $\eta=0.24$, 0.18, 0.12, and 0.024, with $e=0.9$, $ i
= 42^{\circ}$, $\theta = 20^{\circ}$ shown over the whole orbit (left)
and over the phase range 0.9-1.1 (right). The interstellar + nebula
column ($\sim 1\times10^{22}\thinspace \rm{cm}^{-2}$) are additional
to these plots.}
\label{fig:results_col}
\end{center}
\end{figure*}

With $\dot{M}_{1} = 3.5\times10^{-3} \rm{M}_{\odot}\yr^{-1}$ (and
$\eta = 0.024$) the peak column density at periastron increases to
$\simeq 2 \times 10^{25} \cm^{-2}$ (Fig.~\ref{fig:results_col}). The
X-ray minimum is flat and has approximately the correct width
($\delta\phi \sim 0.03$, corresponding to 60 days), yet, the decline
to, and egress out of, minimum are too shallow, and the rest of the
lightcurve is a poor match to the observations. If increased mass-loss
is responsible for the flat bottom of the minimum,
Fig.~\ref{fig:results_lc_eta} shows that it must be confined to a
short phase interval around periastron \citep[see
also][]{Corcoran:2001}. The effect of a short-lived episode of
increased mass-loss is discussed further in \S~\ref{sec:minimum}.

\begin{figure*}
\begin{center}
    \begin{tabular}{ll}
      \resizebox{80mm}{!}{{\includegraphics{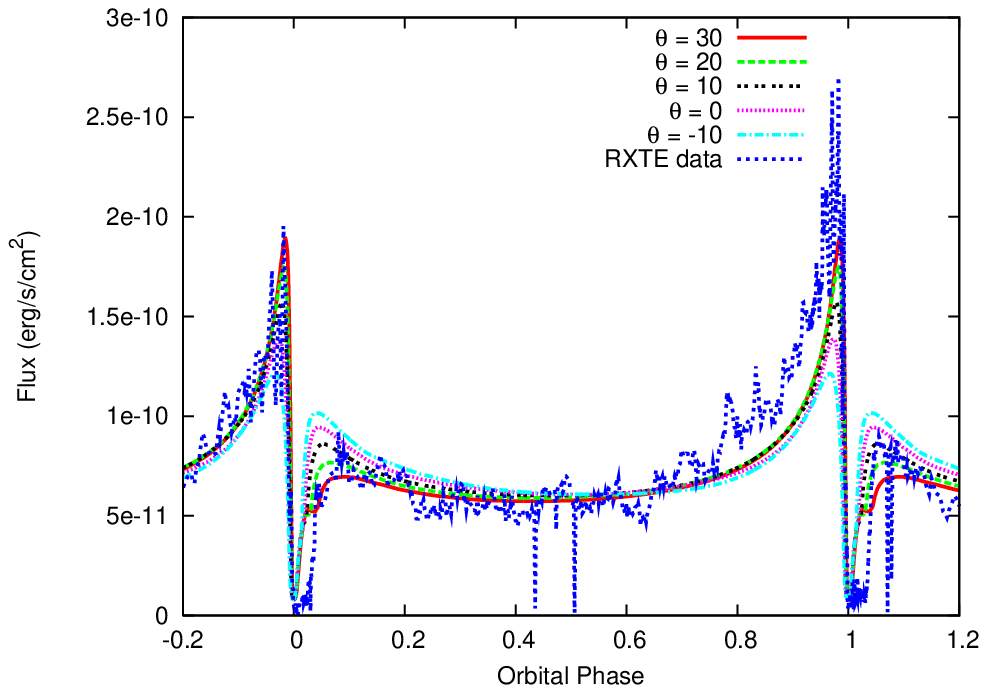}}} &
\resizebox{80mm}{!}{{\includegraphics{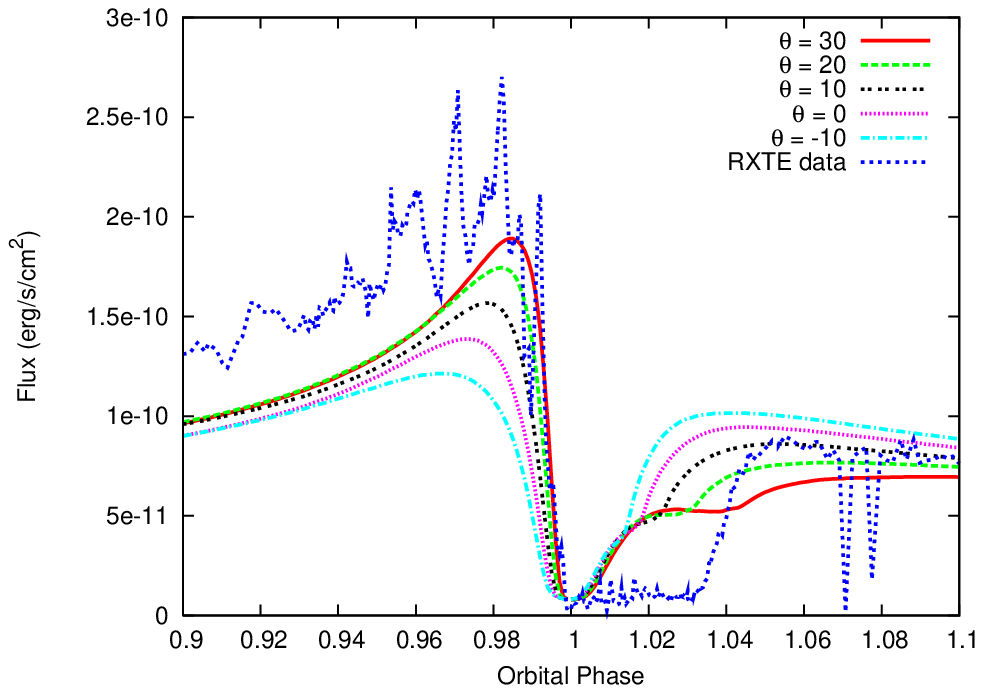}}} \\
    \end{tabular} 
\caption[]{Variation of the attenuated 2-10 keV X-ray emission for
  $\eta = 0.18$, $e = 0.9$, $ i = 42^{\circ}$, and values of $\theta $
  of $-10^{\circ}$ to $30^{\circ}$ in increments of $10^{\circ}$ over
  the whole orbit (left) and over the orbital phase range 0.90-1.10
  (right). The orbital and wind parameters used are noted in
  Tables~\ref{tab:testparameters} and \ref{tab:results}. The model
  results are plotted against the second orbit \textit{RXTE} data
  (which has a higher pre-minimum flux than the first orbit data, see
  e.g. Figs.~\ref{fig:rxtecurve} and \ref{fig:wcrcollapse}) for
  comparison.}
\label{fig:results_lc_los}
\end{center}
\end{figure*}

\subsection{Comparison to the X-ray data}
\label{subsec:fittingthedata}

In the following subsections we examine fits to the \textit{RXTE}
lightcurve and hardness ratio, and \textit{XMM-Newton} spectra.

\subsubsection{Fits to the \textit{RXTE} lightcurve}
\label{subsec:rxtefits}

Having explored how the lightcurves vary with $i, \theta, e$ and
$\eta$, we performed further simulations to obtain a best fit to the
\textit{RXTE} data. The binary models of \cite{Abraham:2005a} and
\cite{Henley:2008} suggest $i \simeq 90^{\circ}$ and $70^{\circ}$
respectively. However, for such values we find a strong disagreement
between the \textit{RXTE} and synthetic lightcurves, mainly due to the
large difference between the pre- and post-minimum X-ray luminosities
in the model lightcurves. A comparison of the $\theta = 180^{\circ}$
curve in Fig.~\ref{fig:los_lc} and the \textit{RXTE} data shows that
an orientation where the secondary star is in front of the primary
during periastron passage \citep{Falceta:2005, Kashi:2007} is not
supported. \cite{Smith:2004} suggested that $\theta =
90^{\circ}$. However, the required pre-/post-minimum flux ratio, and
duration of the minimum, cannot be obtained with any sensible
inclination angle when $\theta=90^{\circ}$, as shown in
Fig.~\ref{fig:los_lc_theta90}. Instead, the \textit{RXTE} data
requires the observer to be situated close to the semi-major axis with
$\theta\simeq 0^{\circ}$ \citep[e.g.][]{Damineli:1996, Pittard:1998,
Corcoran:2001, Corcoran:2005, Akashi:2006, Henley:2008,
Hamaguchi:2007, Nielsen:2007, Damineli:2008a}, whereby the companion
star moves behind the primary at periastron or just after. A detailed
analysis of parameter space using ``chi-by-eye'' yields best-fit
values of $i \simeq 42^{\circ}$ and $\theta \simeq 20^{\circ}$, in
good agreement with \cite{Okazaki:2008}. 

A good match to most of the lightcurve is obtained with $i =
42^{\circ}$, $\theta = 0 - 30^{\circ}$, $\eta=0.18$ and $e=0.9$ (see
Fig.~\ref{fig:results_lc_los}), though a flat extended minimum cannot
be produced. Interestingly, \cite{Corcoran:2005} and
\cite{Hamaguchi:2007} discussed two phases to the minimum; an initial
deep minimum and a shallower phase which begins approximately halfway
through. This behaviour is seen in the model lightcurves and can be
understood by examining Figs.~\ref{fig:abs_col} and
\ref{fig:results_col}, which show a step in the emission weighted
column density due to absorption by the cooled postshock primary wind
exceeding the absorption through the unshocked winds. Note that this
step is not visible in the $\eta = 0.024$ lightcurve in
Fig.~\ref{fig:results_lc_eta} as the attenuation due to the unshocked
winds always dominates.

The similarity between the model and \textit{RXTE} lightcurves with
$\eta = 0.18$ shows that the system can be described well for the
majority of the orbit with one set of parameters. However, the exact
shape of the minimum could not be matched with any single set of
mass-loss rates and terminal wind speeds. In particular, the
absorption peak (Fig.~\ref{fig:results_col}) does not have the
necessary duration to create a flattish minimum. The steep rise out of
minimum observed by \textit{RXTE} also cannot be replicated. This
indicates a need for some additional physics around periastron passage
which causes the column to stay at $\sim 10^{25}\rm{cm}^{-2}$ until
$\phi \sim 1.03$ (\textit{XMM-Newton} observations find the column to
be $\gtsimm 10^{23}\rm{cm}^{-2}$ over this phase range) or some
physical process to reduce the intrinsic emission. A number of
mechanisms which could effect the structure of the WCR and the
observed emission are discussed in \S~\ref{sec:minimum}.

\begin{figure*}
\begin{center}
    \begin{tabular}{l}
      \resizebox{160mm}{!}{{\includegraphics{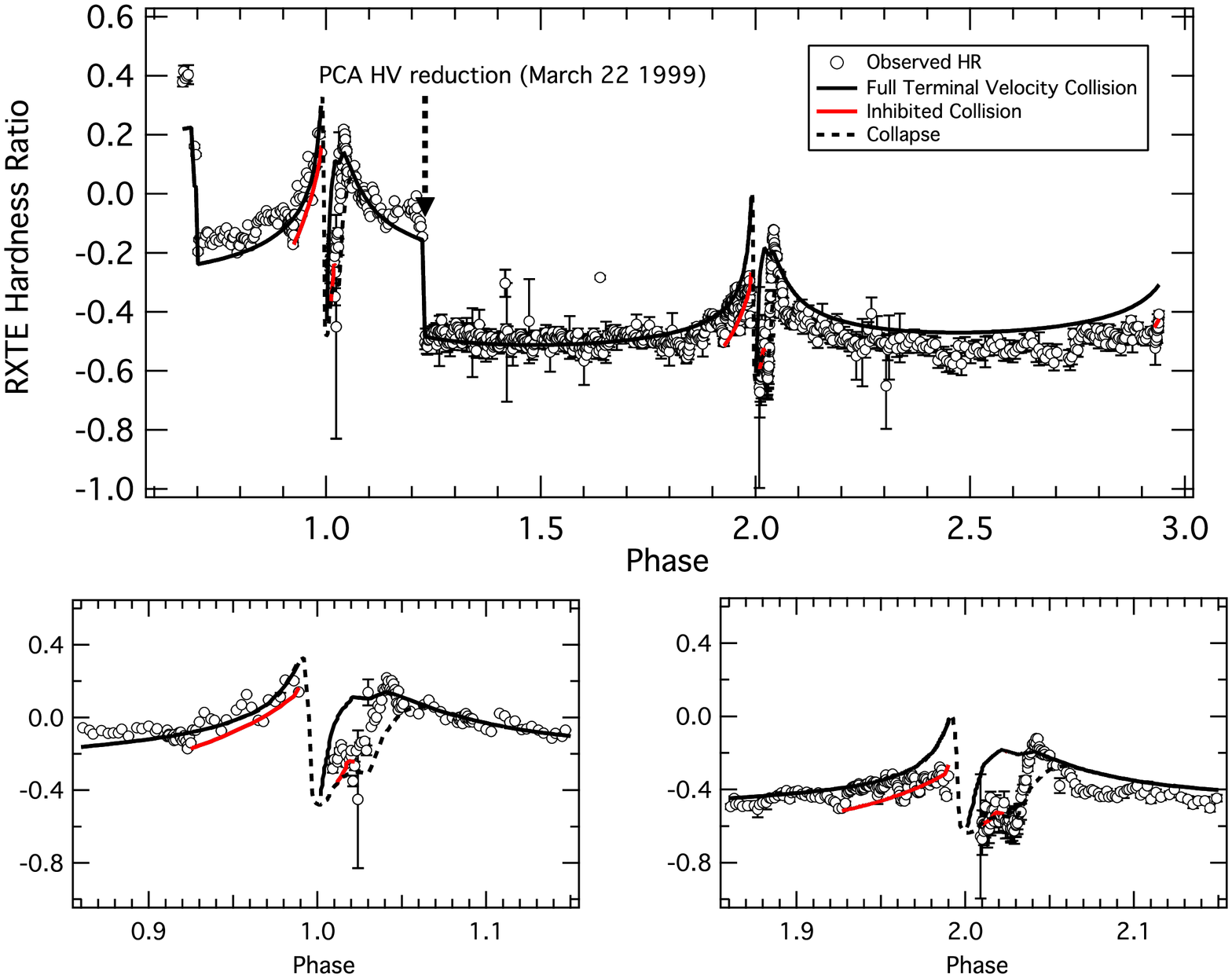}}}
      \\
    \end{tabular} 
\caption[]{Variation of hardness ratio with phase for the best fit
  $\eta = 0.18$ model and the \textit{RXTE} data over the 2.5
  orbits. The model hardness ratio has had the detector response
  folded in for each specific observation. Due to fluctuations in the
  calibrated channel energy gains there is some jagged structure to
  model hardness ratio leading up to periastron passage. The large
  drop at $\phi = 1.23$ corresponds to a gain reduction in the PCU
  on-board \textit{RXTE}, and is also incorporated in the folded
  model. The change at $\phi$ = 0.7 is due to a change in the detector
  response model. The soft band ($s$) is 2-5 keV, the hard band ($h$)
  is 7-10 keV, and the ratio is calculated as $hr=(h-s)/(s+h)$. The
  dip in the model ratio at periastron is due to occultation of the
  WCR by the primary star. Over the orbital phase range $0.924 \ltsimm
  \phi \ltsimm 1.023$ and $1.924 \ltsimm \phi \ltsimm 2.023$ the
  hardness ratio values are from the individual models with lower
  companion wind velocities, discussed in \S~\ref{subsec:rad_braking},
  are shown (red). In the $2.0\ltsimm \phi \gtsimm 3.0$ cycle the
  model appears softer than in the previous cycle, however this is the
  result of processing the model data with the detector response.}
\label{fig:results_hr}
\end{center}
\end{figure*}

Since the intrinsic X-ray luminosity scales as $L_{\rm{X}} \propto
\dot{M}^{2}$ in our model (where it is assumed that the postshock
secondary wind is adiabatic), the normalization of the lightcurve can
indicate the mass-loss rates of the stars. We find that when model
lightcurves produced with the wind parameters in
Table~\ref{tab:testparameters} are compared to the \textit{RXTE} data
the normalization of the model is a factor of $\simeq 2.2$ too low,
which for fixed $\eta$ corresponds to an underestimate in
$\dot{M}_{2}$ of $\simeq 1.5$. Various factors contribute to this
discrepancy, such as the distance to the source and the interstellar
column adopted in the calculations compared to those of PC02, who
assumed a distance to \etacar of 2.1 kpc, and obtained a combined
circumstellar and interstellar column density of $\simeq
7.7\times10^{22}\rm{cm}^{-2}$. In contrast, this work adopts an
interstellar + nebula column of $1\times10^{22}\rm{cm}^{-2}$ (the
majority of which being due to the Homunculus nebula as the ISM column
density is only $\simeq3\times10^{21}\rm{cm}^{-2}$) and a distance of
2.3 kpc \citep[determined from long-slit spectroscopic observations
with the \textit{HST} by][]{Hillier:2001}. In addition, the lack of
spatial resolution with \textit{RXTE} will lead to additional 2-10 keV
flux from nearby, unresolved sources, including the constant
components identified by \cite{Hamaguchi:2007}. In order to compare
like-with-like a background spectrum of these unresolved sources has
been added to all of the model lightcurves where a direct comparison
is made against \textit{RXTE} data, e.g.,
Figs.~\ref{fig:results_lc_los}, ~\ref{fig:results_hr}, and
\ref{fig:wcrcollapse}. Despite this addition, a small discrepancy
remains, and slightly larger $\dot{M}$'s ($\dot{M}_{1} =
4.7\times10^{-4}\Msolpyr$ and $\dot{M}_{2} =
1.4\times10^{-4}\Msolpyr$) than determined by PC02 are required to
bring the model results into agreement with observations.

\subsubsection{The hardness ratio}
\label{subsec:hardratio}

A comparison of the hardness ratio from the best-fit $\eta = 0.18$
model and the \textit{RXTE} data is shown in
Fig.~\ref{fig:results_hr}. We find there is a good agreement between
the model and the \textit{RXTE} data for large parts of the orbit. The
model hardness ratio initially rises towards periastron as stronger
absorption preferentially reduces the soft-band flux. This rise is
also seen in the \textit{RXTE} data. However, the most noticeable
feature in the \textit{RXTE} data is a sudden softening of the flux
which lasts throughout the X-ray minimum. In contrast, the model flux
is hard for the duration of the X-ray minimum, except for a brief
period when the head of the WCR is occulted by the primary star. The
lack of agreement between the data and the model indicates that the
X-ray minimum is not solely caused by absorption and occultation
effects. Instead, there appears to be a substantial intrinsic
reduction of the hardest emission in the \textit{RXTE} data during the
minimum. As we shall see, this also helps to explain the discrepancy
which exists between the lightcurves in
Fig.~\ref{fig:results_lc_los}. The observational data used to
calculate the hardness ratio includes emission from prominent
background sources in the \textit{RXTE} field-of-view such as
W$\thinspace$R 25 and Trumpler 14 and 16. The flux from these sources
is predominantly in the soft band and accounts for the fact that
although the $\phi = 1.009, 1.015,$ and 1.023 spectra from \etacar
shown in Fig.~\ref{fig:results_specs} clearly have higher hard band
flux than soft band, the softening in the \textit{RXTE} data may be
due to imperfect accounting for the soft X-ray background.

An interesting result is that there is a softening in the observed
flux from \etacar after phase $\phi \sim 2.1$, which is highlighted by
the fact that although the \textit{RXTE} data appears very similar
through both cycles there is a noticeable difference between the model
and the data when both have the detector response included. This is
suggestive of an intrinsic variation in the source of the emission,
but uncertainties in the assumed PCU detector response in the 2-10 keV
band cannot yet be ruled out.

\subsubsection{The X-ray spectra}
\label{subsec:spectra}

\begin{figure*}
\begin{center}
    \begin{tabular}{ll}
      \resizebox{75mm}{!}{{\includegraphics{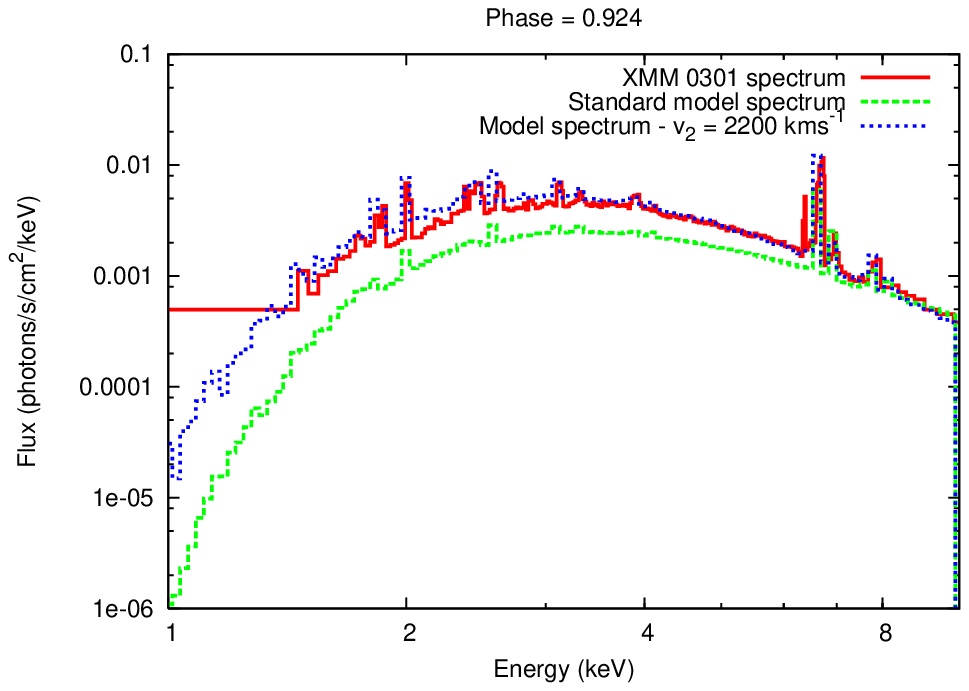}}}
& \resizebox{75mm}{!}{{\includegraphics{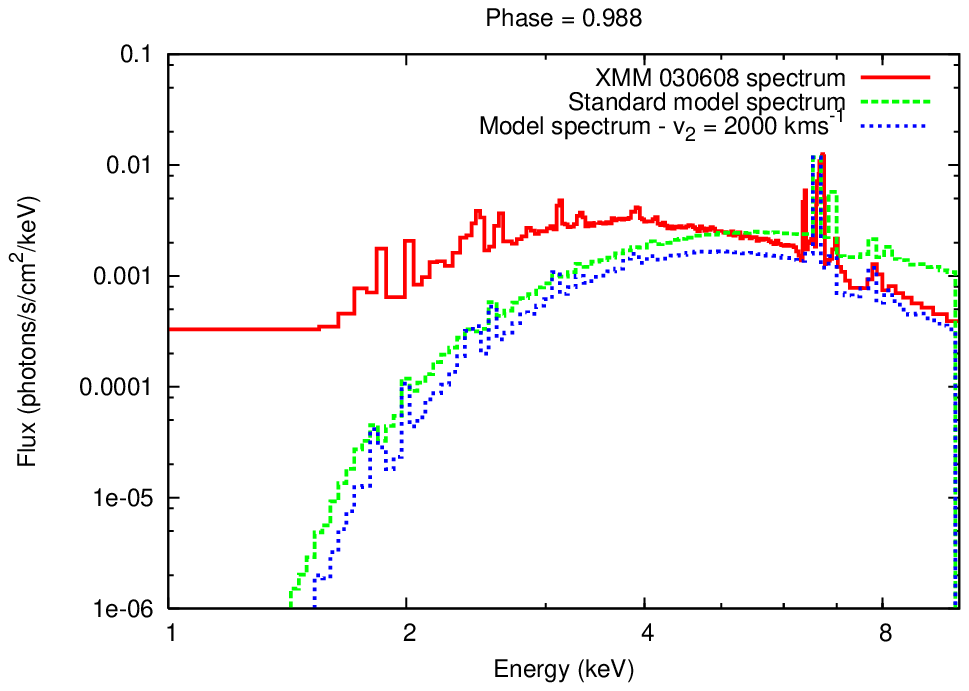}}} \\
\resizebox{75mm}{!}{{\includegraphics{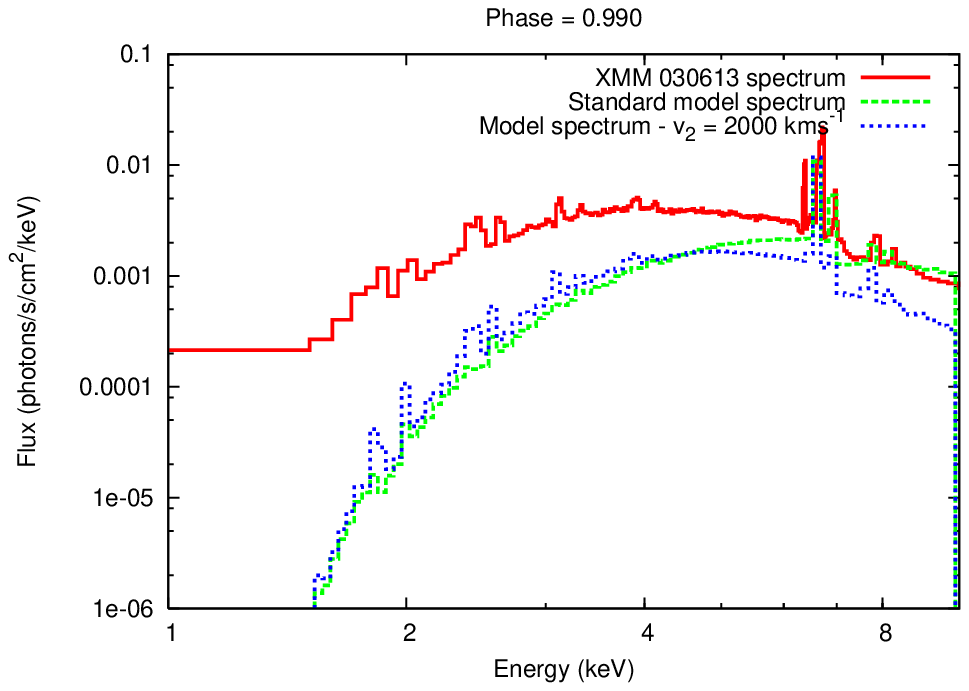}}} &
\resizebox{75mm}{!}{{\includegraphics{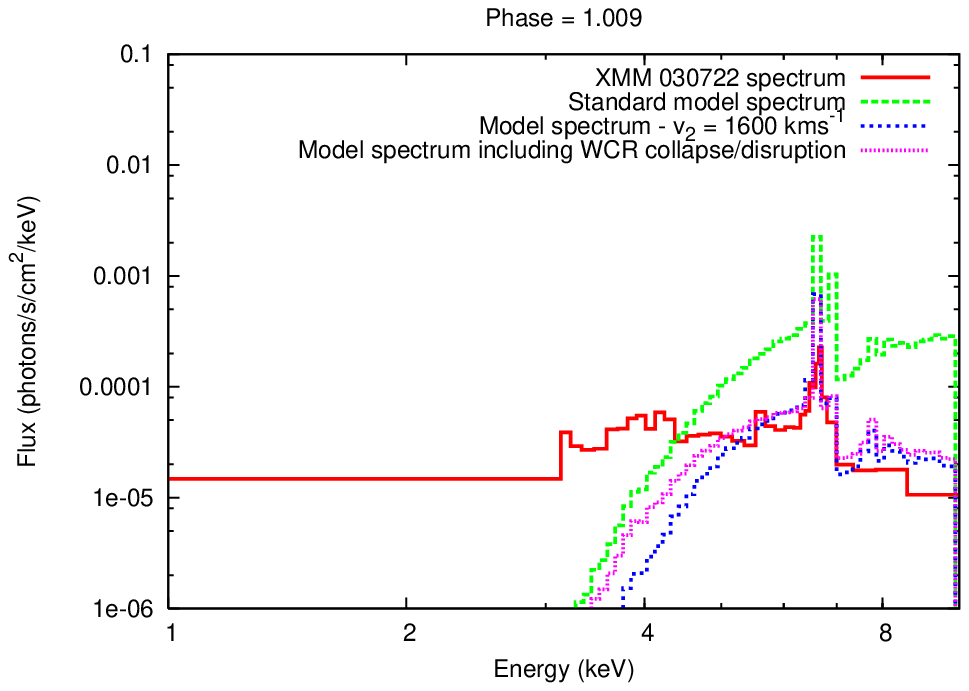}}} \\
\resizebox{75mm}{!}{{\includegraphics{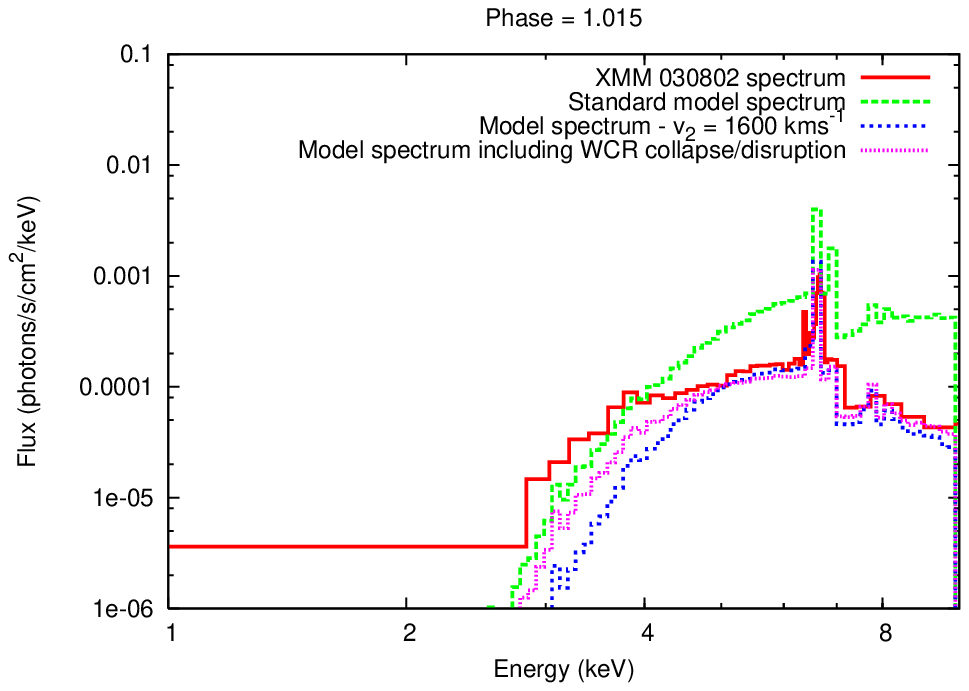}}} &
\resizebox{75mm}{!}{{\includegraphics{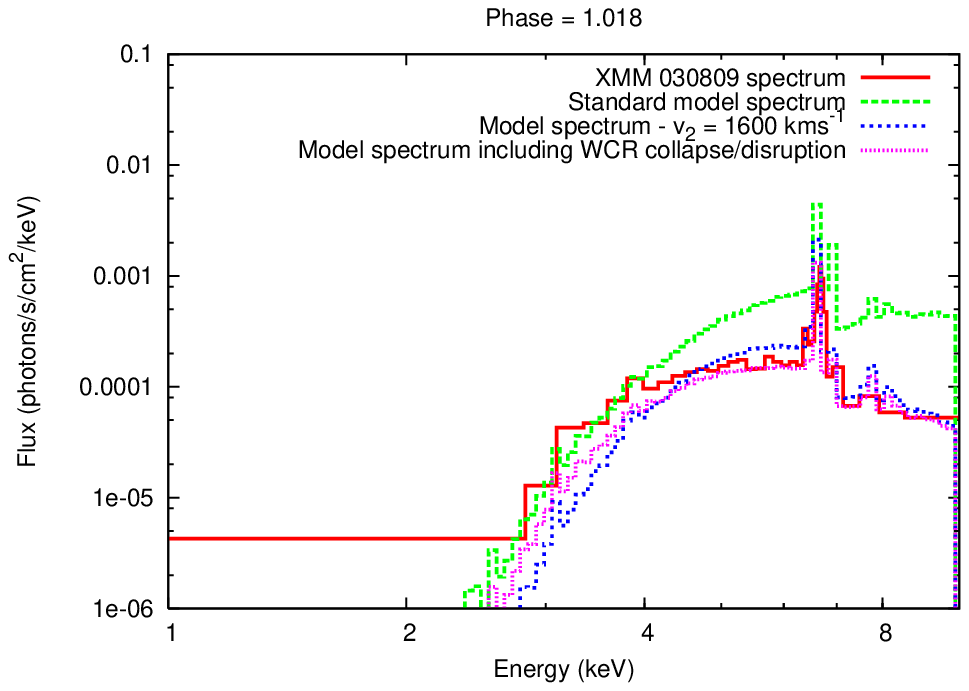}}} \\
\resizebox{75mm}{!}{{\includegraphics{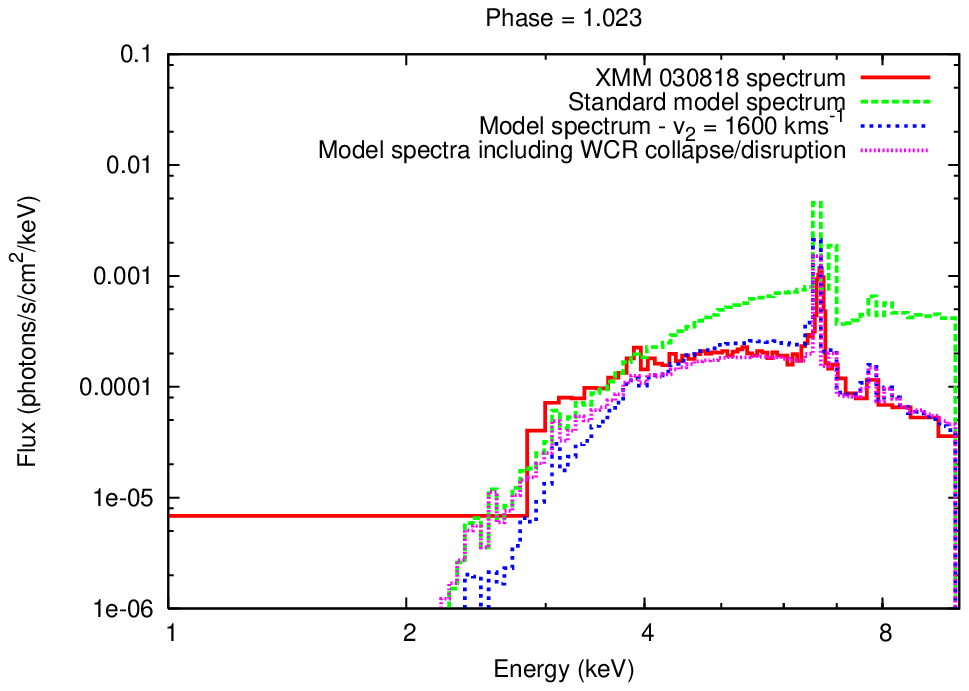}}} & \\

    \end{tabular} 
\caption[]{1-10 keV model spectra with $\eta = 0.18$, $i = 42^{\circ}$
and $\theta = 20^{\circ}$ plotted against \textit{XMM-Newton} data
\citep[see ][]{Hamaguchi:2007} as a function of phase. In the
`standard' model the companions wind is assumed to have a preshock
speed of $3000 \;\rm{km\thinspace s}^{-1}$. Results are also shown for
models with a reduced preshock companion wind velocity. A model which
includes a collapse/disruption of the WCR is also included in the
panels at phases $\phi = 1.009, 1.015, 1.018$, and $1.023$. Models
with lower wind speeds or a collapse of the WCR match the observed
flux above 5 keV during the minimum much better.}
\label{fig:results_specs}
\end{center}
\end{figure*}

A series of \textit{XMM-Newton} observations obtained through the 2003
periastron passage have been analyzed by \cite{Hamaguchi:2007}. The
observed spectra are a combination of a variable hard component
originating from the WCR, and a number of other non-variable
components. The \textit{XMM-Newton} spectra in
Fig.~\ref{fig:results_specs} have the non-variable emission components
from the outer ejecta, the X-ray Homunculus nebula, and the central
constant emission component identified by \cite{Hamaguchi:2007}
removed. These spectra are compared to those produced from our best
model fit to the \textit{RXTE} lightcurve. The Fe line at $\sim$ 7 keV
has a large fluorescence component which is not modelled in the
synthetic spectra; the goal of this work is to reproduce the
broad-band spectra, and not the lines. Recently, \cite{Henley:2008}
examined the variability of the Sulfur and Silicon emission to study
the flow dynamics within the WCR. Future work will focus on developing
the dynamic model to perform individual studies of spectral lines.

At $\phi = 0.924$ we find that there is a significant deficiency in
the emission from our `standard' model (with $v_{\infty2} =
3000\;\rm{km\thinspace s}^{-1}$) at the softer end of the spectrum ($E
= 1-4\;$keV). Thus the model spectrum is slightly harder than the
data, indicating that the postshock temperature of the companion's
wind is over-estimated. The temperature reduction in the postshock
companion wind could be partly caused by mixing of the cold postshock
primary wind into the hot postshock secondary wind, which softens the
spectrum, or radiative inhibition (see
\S~\ref{subsec:rad_braking}). An alternative explanation is that
efficient particle acceleration causes a weaker sub-shock
\citep[c.f.][]{Pittard:2006}. Strong evidence for particle
acceleration in \etacar has recently been presented by
\cite{Leyder:2008}, who detected a hard X-ray tail ($E \sim 10-30
\rm{keV}$). Including the effects of efficient particle acceleration
in the models used in this paper is beyond the scope of the current
work. The $\phi = 0.988$ spectrum occurs at the onset of the
minimum. Although the stars are close at this point, it is still
unlikely that the WCR has entered the acceleration region of the
companion's wind ($d_{\rm{sep}} \simeq 3.7\thinspace \rm{au}$ for
$a=16.64\;\rm{au}$ and $e=0.9$ c.f. Fig.~\ref{fig:radinh_vels}).

The quasi-periodic ``flares'' observed in the \textit{RXTE} lightcurve
\citep{Corcoran:2005} are not reproduced in the synthetic data. These
may result from dynamic instabilities in the WCR or stellar winds and
cannot be reproduced in the dynamic model used in this work which is
assumed to have a stable WCR. The \textit{XMM-Newton} observations at
$\phi=0.988$ and 0.990 occur at the bottom and top of a flare peak
respectively, which accounts for some of the discrepancy seen between
the data and model spectra.

The spectra from the `standard' model at $\phi = 1.009, 1.015,
1.018,\thinspace \rm{and} \; 1.023$ also show significant excess flux
at high energies. 

\section{The nature of the X-ray minimum}
\label{sec:minimum}

\begin{figure}
\begin{center}
\psfig{figure=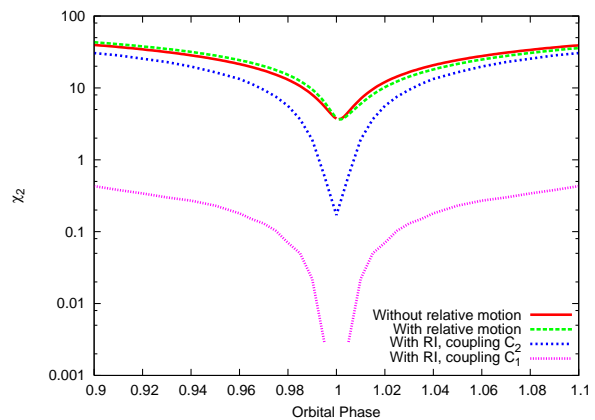,width=8.0cm}
\caption[]{Variation of the cooling parameter, $\chi_{2}$, for the
companion wind in the orbital phase range $\phi = 0.9 - 1.1$. Results
are shown for calculations with (green) and without (red) the
inclusion of the relative motion of the stars, which affects the
position of the ram pressure balance and the preshock wind
speeds. Prior to periastron the stars approach each other, and the
relative wind speeds are higher, increasing $\chi$ compared to the
static case. The opposite effect occurs after periastron. Both curves
reach a minimum at periastron. For simplicity the skew of the WCR due
to orbital motion is neglected, and the wind is assumed to be
perfectly smooth. Clumping and efficient particle acceleration
\citep[see e.g.][]{Pittard:2006} will reduce $\chi_{2}$ below the
level shown. The effect on $\chi_{2}$ due to radiative inhibition is
also shown (see \S~\ref{subsec:rad_braking}).}
\label{fig:cooling}
\end{center}
\end{figure}

In the previous section model simulations were performed which
provided a reasonable fit to the lightcurve as a whole but did not
give a good fit to the minimum. A detailed examination of the spectrum
and hardness ratio around the minimum showed some significant
discrepancies between the model results and the observations, with the
model emission generally harder than the data during the minimum. In
this section possible solutions for resolving this discrepancy are
discussed.

A plausible scenario is that around periastron the preshock speed of
the secondary wind is reduced to the point that it does not shock to
high enough temperatures to produce the hard X-ray emission which is
normally seen. This change in the preshock velocity may be initiated
by the WCR moving into the acceleration zone of the secondary wind due
to the high orbital eccentricity, or to a period of enhanced mass-loss
from the primary star. Alternatively the acceleration of the secondary
wind may be inhibited towards the primary star by its enormous
radiation field. In addition, the shocked secondary wind may become
radiative around periastron (especially if radiative inhibition is a
strong effect). If both winds strongly cool, it is possible that the
WCR disrupts and breaks up into a mass of cold dense blobs, each
surrounded by oblique shocks as they are impacted by the supersonic
winds (i.e. the WCR ``discombobulates'' - Kris Davidson, personal
communication). The fast companion wind may then be slowed through a
sequence of oblique, radiative shocks, each with a lower postshock
temperature than would be obtained behind a single global shock.

\begin{figure*}
\begin{center}
    \begin{tabular}{ll}
      \resizebox{80mm}{!}{{\includegraphics{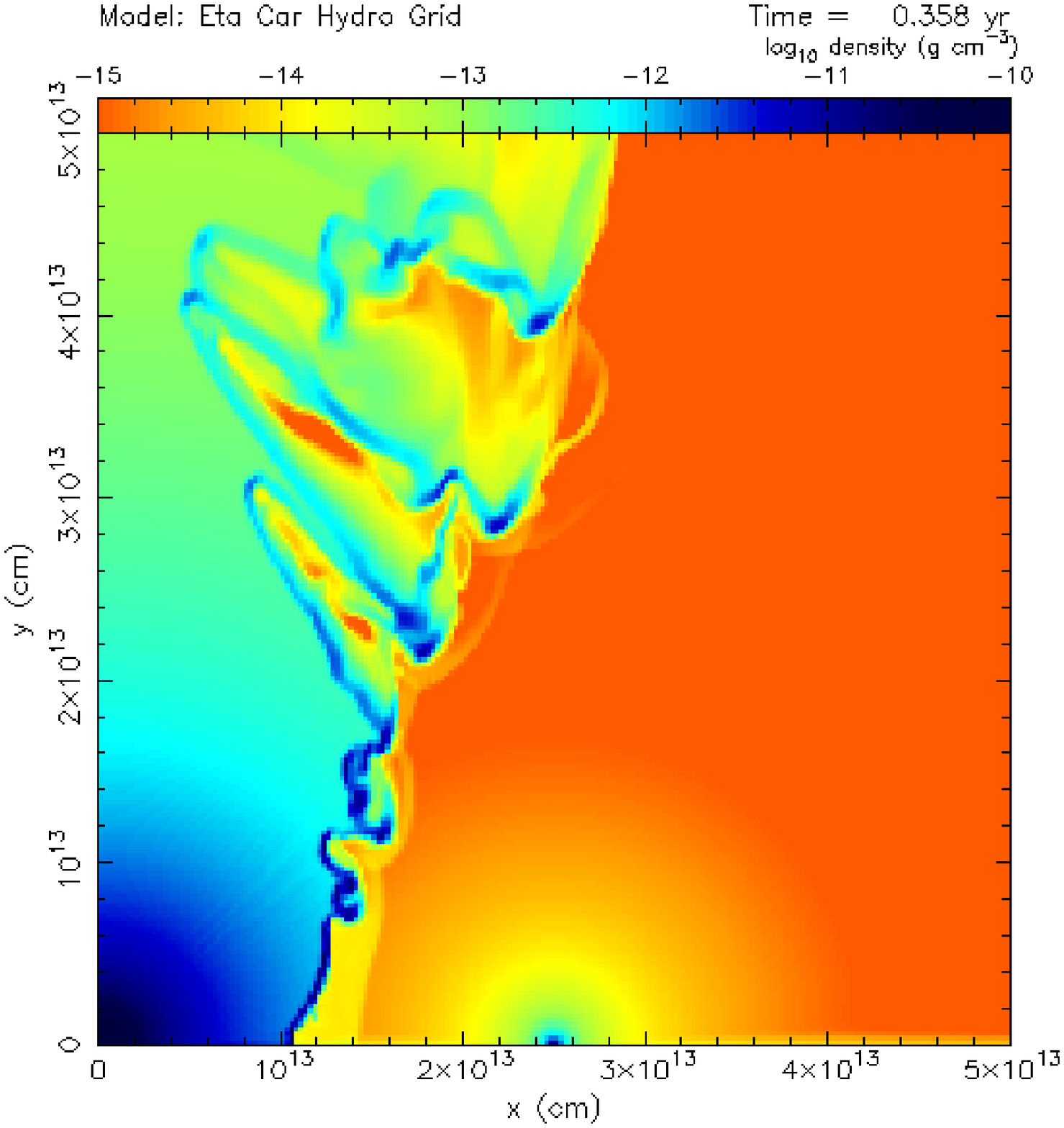}}} &
\resizebox{80mm}{!}{{\includegraphics{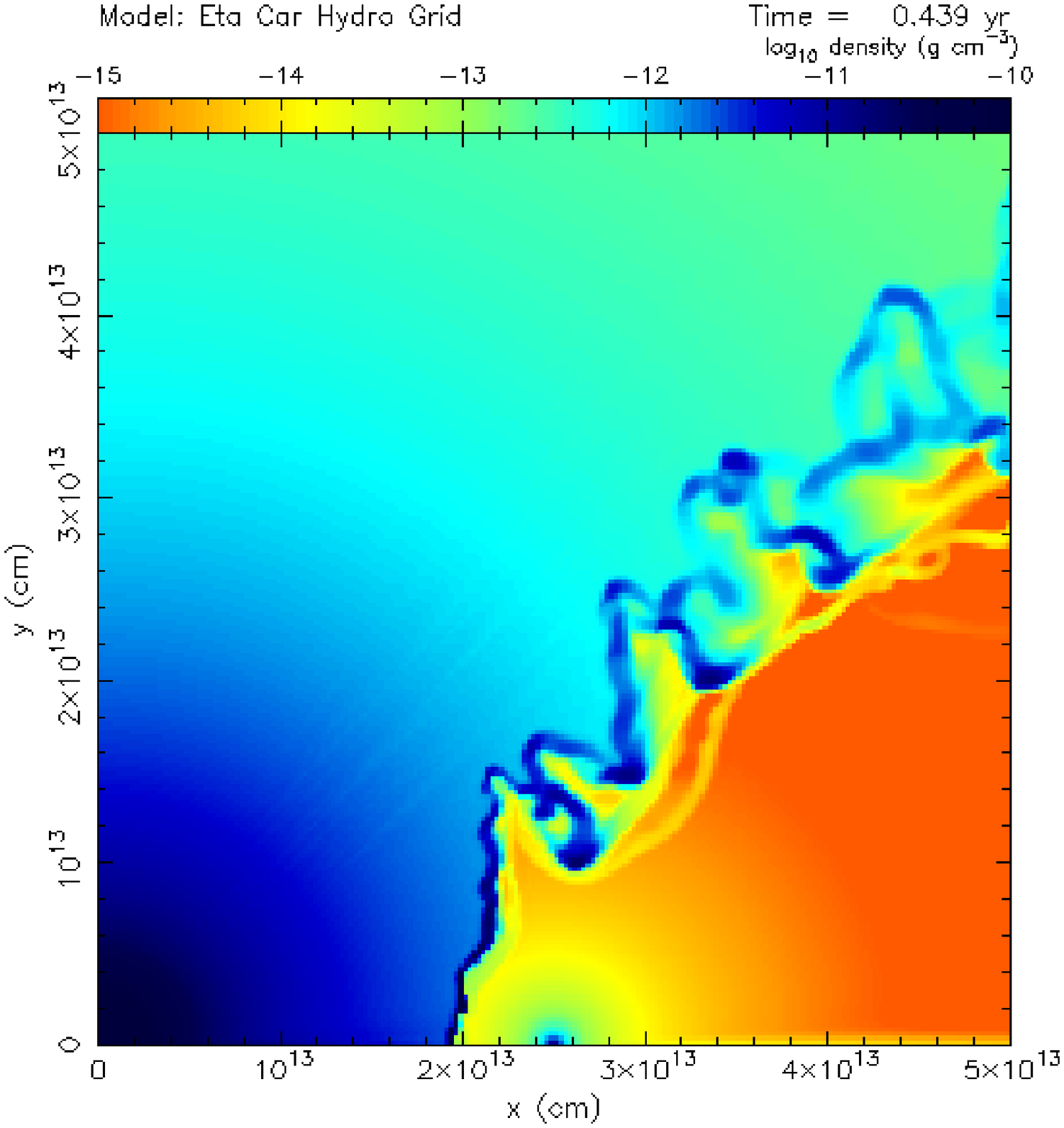}}} \\
    \end{tabular} 
\caption[]{2D hydrodynamic simulations of the WCR at a separation
corresponding to periastron for $e=0.9$. The plots show the effect of
increasing the primary mass-loss rate ($\dot{M}_{1} = 5.0
\times10^{-4}$ (left panel) and $3\times10^{-3}$ (right panel)
\Msolpyr). In each simulation the wind speeds are set according to the
on-axis preshock speed determined from a simple consideration of
momentum balance (see Table~\ref{tab:hydro_cooling} where the
effective wind momentum ratio along the line-of-centres is also
given). As the WCR is pushed into the acceleration zone of the
secondary wind the shocked secondary wind becomes increasingly
radiative ($\chi_{2} =$ 3.6 in the left panel and 0.71 in the right
panel) and the WCR becomes increasingly unstable. No gravitational or
radiative driving forces are included in these calculations.}
\label{fig:hydro_cooling}
\end{center}
\end{figure*}

However, the above processes by themselves cannot account for the long
duration of the minimum, and the real picture is likely to be even
more complicated. There is a possibility that the WCR collapses onto
the surface of the secondary star during the close approach of the
stars around periastron. This collapse would be initiated by the winds
failing to attain a stable momentum balance at some
point. Hydrodynamic simulations of a collapse of the WCR in the
$\iota$ Orionis system \citep{Pittard:1998b} lead to a minimum in the
flux and temperature of the X-ray emission at periastron
\citep{Pittard:2000}. A similar event may occur in the \etacar
system. In both the ``collapse'' and ``discombobulation'' scenarios
mentioned above, it may take some considerable time to re-establish a
bona-fide WCR.

\subsection{Effects of wind acceleration and variable mass loss}
\label{subsec:cooling_effects}

The models in \S~\ref{sec:results} have been made with the assumption
that the winds collide at terminal velocity. In reality it is likely
that one, or both, of the winds may still be accelerating when they
collide. To gauge how this affects the previous results, the position
of the stagnation point has been computed using $\beta$-velocity laws
for both winds, with $\beta$ = 4 and 1 for the primary and secondary
winds respectively. Compared to the terminal winds case, ram pressure
balance occurs closer to the primary star at all orbital phases (since
its wind accelerates more slowly). For instance, at $\phi = 1.0$ the
preshock primary wind only reaches a velocity of $v_{1} =
125\;\rm{km\thinspace s}^{-1}$ and ram pressure balance occurs at a
distance of $0.44\thinspace d_{\rm{sep}}$ from the companion star,
compared to $0.30\thinspace d_{\rm{sep}}$ when terminal velocity winds
are assumed. Cooling parameters for both winds were then
calculated. As expected, the primary's wind was found to be highly
radiative throughout the orbit ($\chi \ll 1$). Interestingly, the
cooling parameter of the shocked companion wind at periastron is at
roughly the point that cooling starts to become non-negligible ($\chi
\sim 3-5$ see Fig.~\ref{fig:cooling}).

\begin{figure}
\begin{center}
    \begin{tabular}{l}
      \resizebox{80mm}{!}{{\includegraphics{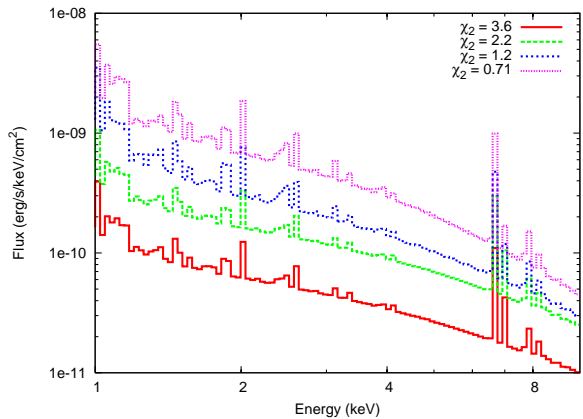}}} \\
    \end{tabular} 
\caption[]{1-10 keV spectra showing the variation in the intrinsic
emission calculated from a 2D hydrodynamic simulation as the mass-loss
rate of the primary star increases. The spectra are labelled by their
respective values of $\chi_{2}$, as noted in
Table~\ref{tab:hydro_cooling}.}
\label{fig:cooling_spec}
\end{center}
\end{figure}

Cooling of the shocked companion's wind becomes more important if
there is a period of enhanced mass-loss from the primary wind (see
Table~\ref{tab:hydro_cooling}) since the WCR is pushed deeper into the
secondary wind resulting in lower preshock velocities and higher
densities (see Fig.~\ref{fig:hydro_cooling}). Note the dramatic change
in the width of the postshock secondary wind (i.e. the distance
between the shock and the contact discontinuity) on the
line-of-centres as $\chi_{2}$ becomes smaller. The intrinsic X-ray
emission from these models is shown in Fig.~\ref{fig:cooling_spec},
where two trends are noticeable. First, the total flux increases with
the primary mass-loss rate since a greater portion of the secondary
wind is shocked, and second, the emission also becomes softer, since
the preshock secondary wind speed declines. It is clear that the net
effect of increasing $\dot{M}_{1}$ is to increase the
\textit{intrinsic} X-ray emission, though this would only be a
temporary effect if any increase in $\dot{M}_{1}$ was short-lived. Any
increase in $\dot{M}_{1}$ cannot last too long as otherwise it would
lead to an observed increase in the X-ray flux as the lines-of-sight
to the emitting plasma exit the primary wind.

\begin{table}
\begin{center}
  \caption[]{The effect of increasing the primary star mass-loss rate
  on the preshock wind speeds ($v_{1}$ and $v_{2}$), the effective
  momentum ratio along the line-of-centres ($\eta_{\rm{loc}}$), the
  distance of the stagnation point from the centre of the companion
  star ($r_{2}$), and the cooling parameter of the secondary wind
  ($\chi_2$). $v_{1}$ and $v_{2}$ are calculated using
  $\beta$-velocity laws ($v_{\infty1}=500\;\rm{km\thinspace s}^{-1}$,
  $v_{\infty2}=3000\;\rm{km\thinspace s}^{-1}$, $\beta_{1} = 4$, and
  $\beta_{2} = 1$). $\dot{M}_{2}= 1.4\times10^{-5}\Msolpyr$ in all
  models (as determined in \S~\ref{subsec:rxtefits}) and the assumed
  stellar separation is $d_{\rm{sep}}= 359\Rsol$ (corresponding to
  periastron in our standard model with $e=0.9$). The radius of the
  gravitationally bound core of the primary star is taken to be
  100\Rsol. $\eta_{\rm{loc}}$ and $\chi_2$ are evaluated using the
  preshock, rather than the terminal, wind speeds along the
  line-of-centres. The orbital aberration of the WCR is neglected for
  simplicity.}
\begin{tabular}{llllll}
\hline $\dot{M_1}$ & $v_{1}$ & $v_{2}$ & $\eta_{\rm{loc}}$ & $r_{2}$ & $\chi_{2}$ \\ 
(\Msolpyr) & (km\thinspace s$^{-1})$ & (km\thinspace s$^{-1}$) & & ($d_{\rm{sep}}$) & \\ 
\hline 
$5.0\times10^{-4}$ & 127 & 2610 & 0.58 & 0.43 & 3.6 \\
$1.0\times10^{-3}$ & 147 & 2490 & 0.24 & 0.33 & 2.2 \\
$2.0\times10^{-3}$ & 164 & 2295 & 0.10 & 0.24 & 1.2 \\
$3.0\times10^{-3}$ & 173 & 2130 & 0.063 & 0.20 & 0.71 \\
\hline
\end{tabular}
\label{tab:hydro_cooling}
\end{center}
\end{table}

\subsection{Radiative braking/inhibition}
\label{subsec:rad_braking}

\begin{table}
\begin{center}
  \caption[]{The parameters used to calculate the line driving of the
  stellar winds, and the effect of radiative
  braking/inhibition. $T_{\rm{cs}}$ is the temperature at the surface
  of the gravitationally bound core of the star, which for the primary
  star is taken to be at a radius of $100\Rsol$. $k$ and $\alpha$ are
  the \cite{Castor:1975} line driving parameters, where subscripts 1
  and 2 are used to define the coupling ($C$) between the winds and
  the radiation fields of the primary and secondary stars
  respectively.}
\begin{tabular}{lll}
\hline
 & Primary  & Secondary \\
\hline
$T_{\rm{cs}}$ (K)& 25,800 & 30,000 \\
$L_{\ast} $ ($10^{6} {\rm L_{\odot}}$) & $4$ & $0.3$  \\
$k$ & 0.351 &  0.443 \\
$\alpha$ & 0.525 & 0.712 \\
$\dot{M}\;(\Msolpyr)$ & $5.0\times10^{-4}$ & $1.4\times10^{-5}$ \\
$v_{\infty}\;(\kmps)$ &  500 & 3000 \\
\hline

\end{tabular}
\label{tab:cakparameters}
\end{center}
\end{table}

\begin{figure}
\begin{center}
    \begin{tabular}{l}
      \resizebox{80mm}{!}{{\includegraphics{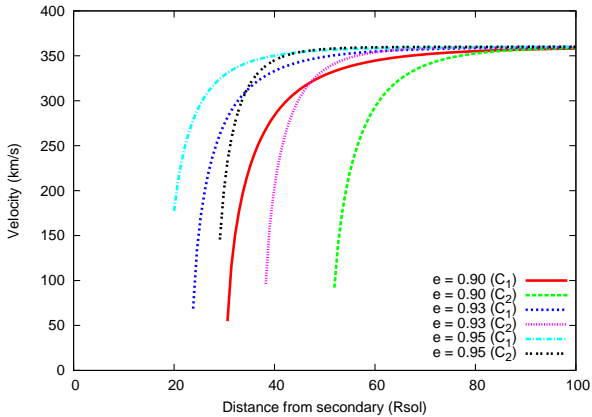}}} \\
    \end{tabular} 
\caption[]{Radiatively braked velocity profile of the primary star's
wind as it approaches the companion star for various orbital
eccentricities and coupling parameters. Radiative braking causes the
pronounced sudden reduction in the velocity of the incoming wind (from
the right in this plot). Orbital motion is not included. The
parameters used in the line driving calculations are noted in
Table~\ref{tab:cakparameters}.}
\label{fig:radbrake}
\end{center}
\end{figure}

The radiation fields of both stars may affect the position and nature
of the WCR around periastron passage. In this section we consider two
possible effects: radiative braking and radiative inhibition. The
former refers to the deceleration of a stellar wind by the radiative
flux from the opposing star prior to reaching the WCR, whereas the
latter refers to the reduction of the net rate of acceleration of a
stellar wind due to the opposing radiation field. With this in mind it
is useful to note that inhibition occurs close to the star driving the
wind, i.e. in the acceleration region of the wind, and braking occurs
close to the WCR. To determine whether radiative braking is effective
for either star, the equations of \cite{Gayley:1997} have been
evaluated. To perform these calculations the line driving theory of
\cite{Castor:1975} (hereafter CAK) was used to find the values of the
parameters $k$ and $\alpha$ required to produce the desired terminal
velocities and mass-loss rates for each star
(Table~\ref{tab:cakparameters}). While there is some uncertainty to
the appropriate values of $k$ and $\alpha$ these values provide a
reasonable starting point for examining the dynamical interaction
between the radiation fields and the winds in this system. We define
coupling $C_1$ as the the coupling between the primary star radiation
and stellar wind, which is described by $k_{1}$ and $\alpha_{1}$, and
respectively coupling $C_{2}$ constitutes the equivalent for the
companion star. An uncertainty in the use of the equations in
\cite{Gayley:1997} is that it is not clear which values of $k$ and
$\alpha$ should be used when considering the braking of one star's
wind by the radiation field of the other \citep[i.e. should the
primary star's $k$ and $\alpha$ values be used when considering
braking by the secondary star's radiation ? see also][for further
discussion]{Pittard:1998b,St-Louis:2005}. Therefore we have
investigated a number of different possibilities. We find that the
primary star's radiation does not brake the companion's wind,
irrespective of which CAK parameters from
Table~\ref{tab:cakparameters} are used. When the close proximity of
the WCR to the companion star is considered this is not such a
surprising result as the ability of sudden radiative braking to halt a
wind is expected to be effective when the incoming wind is close to
impacting the surface of a star, which is not the case for the
companion star's wind travelling towards the primary star. However,
the opposite scenario of the companion's radiation field braking the
incoming primary star's wind, does occur. Fig.~\ref{fig:radbrake}
shows that in this respect alone at an orbital eccentricity of
$e=0.90$, braking is significant when the coupling is with either star
(i.e. coupling $C_1$ or $C_2$). However, unless the coupling $C_1$ is
used, the braking does not occur before the primary wind has reached
the WCR. The location of sharp braking moves closer towards the
secondary star as $e$ increases, meaning that a normal ram-ram balance
is increasingly likely. It is unclear whether the high density of the
primary's wind will in reality prevent effective braking, as the
argument of \cite{Kashi:2008} does not consider the ram pressure of
the companion wind acting on the dense blobs of primary wind
\textit{and} the subsequent rate of ablation which may destroy the
blobs before they reach the companion star's surface. A more in-depth
examination of this mechanism is beyond the scope of this work, but
will be considered in future models. Our current calculations improve
upon the approach taken by \cite{Kashi:2008} by determining the full
range of coupling parameters.

\begin{figure}
\begin{center}
    \begin{tabular}{l}
      \resizebox{80mm}{!}{{\includegraphics{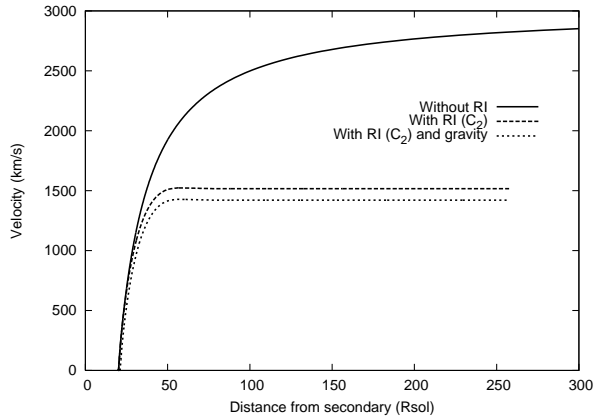}}} \\
    \end{tabular} 
\caption[]{The speed of the companion wind towards the primary star
(located at a distance of 359 \Rsol which corresponds to the
separation of the stars at periastron for $e=0.9$) assuming that the
coupling of the primary star's radiation field to the companion star's
wind is described by coupling $C_2$ (i.e. the same values of the CAK
parameters as required for the companion star to drive its own
wind). Clearly, the acceleration of the wind is strongly inhibited by
the radiation field of the primary star.}
\label{fig:radinh_vels}
\end{center}
\end{figure}

\begin{figure}
\begin{center}
    \begin{tabular}{l}
      \resizebox{80mm}{!}{{\includegraphics{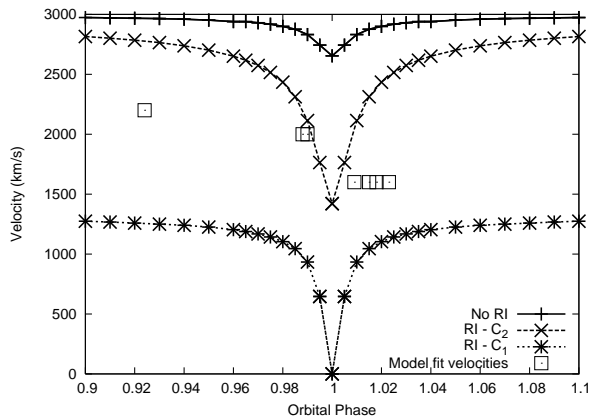}}} \\
    \end{tabular} 
\caption[]{The preshock velocity of the companion's wind as a function
of orbital phase for $e=0.9$. Three scenarios are plotted: no
radiative inhibition (no RI), RI with coupling $C_2$ (see
\S~\ref{subsec:rad_braking}), and RI with coupling $C_1$. The preshock
velocity was calculated after determining the location of the
stagnation point between the winds. The lower curve does not have a
stable balance point at periastron, and the companion cannot launch a
wind towards the primary star. Orbital motion is not included in this
plot. The parameters used in the line driving calculations are noted
in Table~\ref{tab:cakparameters}. The squares represent the preshock
velocities used to attain improved fits to the \textit{XMM-Newton}
spectra in Fig.~\ref{fig:results_specs}.}
\label{fig:radinh_psvels}
\end{center}
\end{figure}

As mentioned above, the radiation fields of the stars can also play a
prominent role in altering the mass-loss rates and terminal velocities
of the stellar winds by inhibiting the acceleration of each other's
wind \citep{Stevens:1994}. When applied to \etacar we find that the
radiation field of the primary star significantly reduces the terminal
velocity of the companion wind (Fig.~\ref{fig:radinh_vels}) to $\simeq
1520\;$km\thinspace s$^{-1}$ (assuming coupling $C_2$). A further
reduction occurs if the gravitational influence of the primary star on
the companion star is considered. In this case tidal deformation
results in the radius of the secondary star towards the primary
increasing by $\sim 1 \Rsol$. This in turn reduces the region
available for wind acceleration, and the terminal velocity and
mass-loss rate of the companion's wind drop down to $\simeq
1420\;$km\thinspace s$^{-1} $ and $1.25\times10^{-5}\Msolpyr$
respectively. When calculations are performed with coupling $C_1$, the
companion's wind speed is significantly reduced at all orbital phases
and the preshock velocity decreases such that a stable momentum
balance cannot be attained at periastron
(Fig.~\ref{fig:radinh_psvels}).

\begin{figure}
\begin{center}
    \begin{tabular}{l}
      \resizebox{80mm}{!}{{\includegraphics{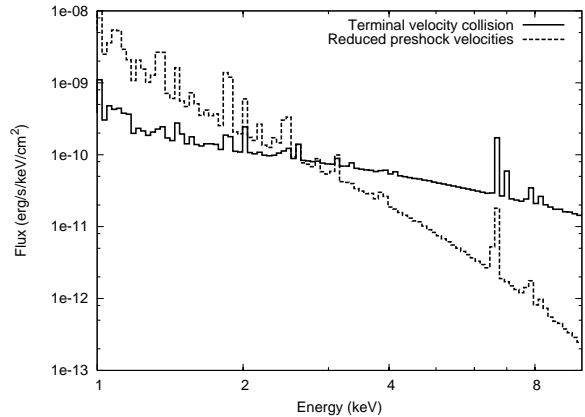}}} \\
    \end{tabular} 
\caption[]{Intrinsic spectra calculated from hydrodynamic simulations
discussed in \S~\ref{subsec:rad_braking} (corresponding to periastron
for $e=0.9$), where the winds are assumed to collide at terminal
velocities (solid) or at much reduced velocities (dashed).}
\label{fig:radinh_specs}
\end{center}
\end{figure}

These alterations to the wind parameters will have drastic
implications for the observed emission and may explain the wind
disturbance discussed by \cite{Martin:2006}. To examine this scenario
we have performed further hydrodynamic simulations where the wind
speeds of the stars are $v_{1} = 360\;\rm{km\thinspace s}^{-1}$ and
$v_{2} = 1420\;\rm{km\thinspace s}^{-1}$ (i.e. accounting for the slow
acceleration of the primary wind, and the radiative inhibition of the
secondary wind with coupling $C_2$). The shocked companion wind
reaches peak temperatures of $<10^{7}\;$K in this case, compared to
$\sim10^{8}\;$K if colliding at $3000\;\rm{km\thinspace s}^{-1}$. The
intrinsic spectra calculated from these hydrodynamic simulations are
shown in Fig.~\ref{fig:radinh_specs}. There is a dramatic reduction in
the emission above 2 keV when the preshock velocity of the companion
is reduced to the value implied by the radiative inhibition
calculations. This provides some explanation for the over-estimation
of the hard band flux in the dynamic model in \S~\ref{sec:results}
where terminal velocity winds are assumed (see
Figs.~\ref{fig:results_hr} and \ref{fig:results_specs}). Models with
reduced preshock companion wind velocities yield much improved fits to
the \textit{XMM-Newton} spectra in Fig.~\ref{fig:results_specs}. For
instance, at $\phi = 0.924$ a better fit can be attained when a
preshock velocity of $2200\;\rm{km\thinspace s}^{-1}$ is used, which
indicates that radiative inhibition is important at this phase. Also,
the excess in high energy flux seen in the `standard' model spectra at
$\phi = 1.009, 1.015, 1.018,\thinspace \rm{and} \; 1.023$ can be
rectified when reduced preshock companion wind velocities are used (a
noteable failing, however, is the discrepency which still exists
between the model and the data at $\phi = 0.988$ and
0.990). Interestingly, the required velocities lie within the range of
predicted preshock velocities caused by radiative inhibition
(Fig.~\ref{fig:radinh_psvels}). Considering the recovery of the
companion wind as the stars recede, we would expect a preshock
velocity of $v_{2} \simeq 2500\;\kmps$ at $\phi\sim 1.13$, which is
significantly higher than the $\sim 1100\;\kmps$ detected by
\cite{Iping:2005} at this phase. However, this comparison does not
account for projected velocity vectors and/or the preshock velocity of
the \textit{downstream} gas. We note that although there is some
uncertainty in the coupling parameters used, the calculations
performed provide a useful guide to the range of inhibited velocities.

It is also useful to consider the degree of radiative inhibition near
apastron. PC02 found a best-fit velocity of $3000\;\rm{km\thinspace
s}^{-1}$, and this value has been adopted in our `standard'
model. Reasonable agreement with this value is attained when using
coupling $C_{2}$, where we find the terminal wind speed is reduced to
$v_{\infty2} = 2916\;\rm{km \thinspace s}^{-1}$. In contrast, with
coupling $C_{1}$, the companion's wind speed is more than halved
giving $v_{\infty2}= 1321\;\rm{km \thinspace s}^{-1}$. Hence, our
calculations support a coupling which is inclined towards the wind
rather than the radiation field in question. At first sight this is in
contrast to the findings of \cite{St-Louis:2005}, where the wind-wind
collision in Sanduleak 1 was investigated, although their conclusion
may depend on the assumed mass-loss rates. It should be noted that in
reality the values of $k$ and $\alpha$ are spatially and phase
dependent \citep[e.g.][]{Puls:2006,Fullerton:2006}, yet to make the
problem tractable we needed to make some simplifying approximations.

Clearly, radiative inhibition is an important mechanism in the \etacar
system, and its effect must be properly considered in future models.

\begin{figure}
\begin{center}
    \begin{tabular}{l}
      \resizebox{80mm}{!}{{\includegraphics{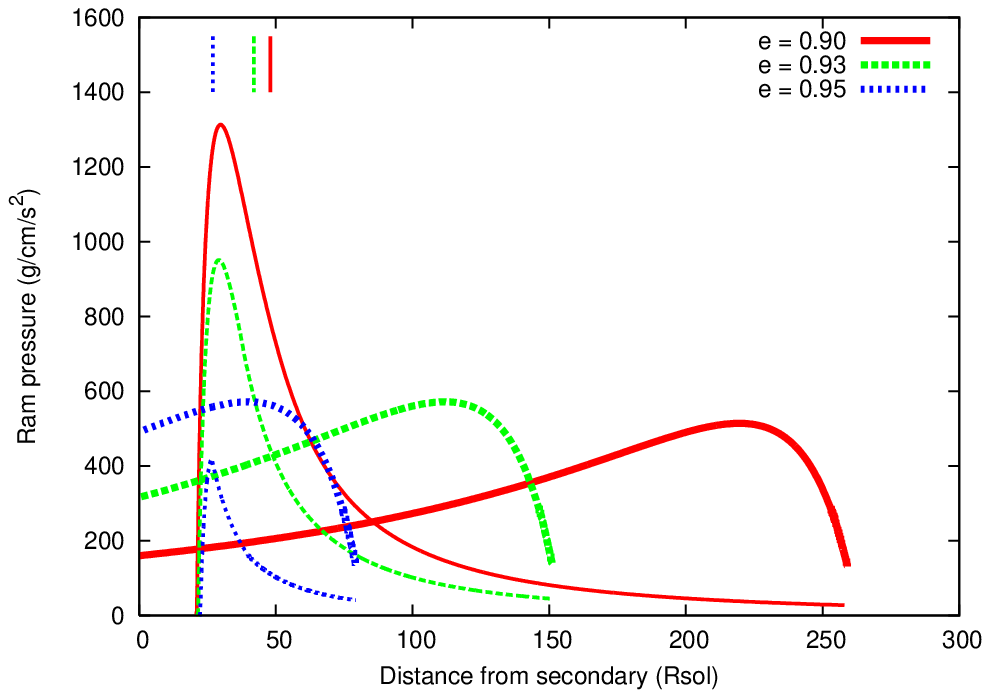}}} \\
      \resizebox{80mm}{!}{{\includegraphics{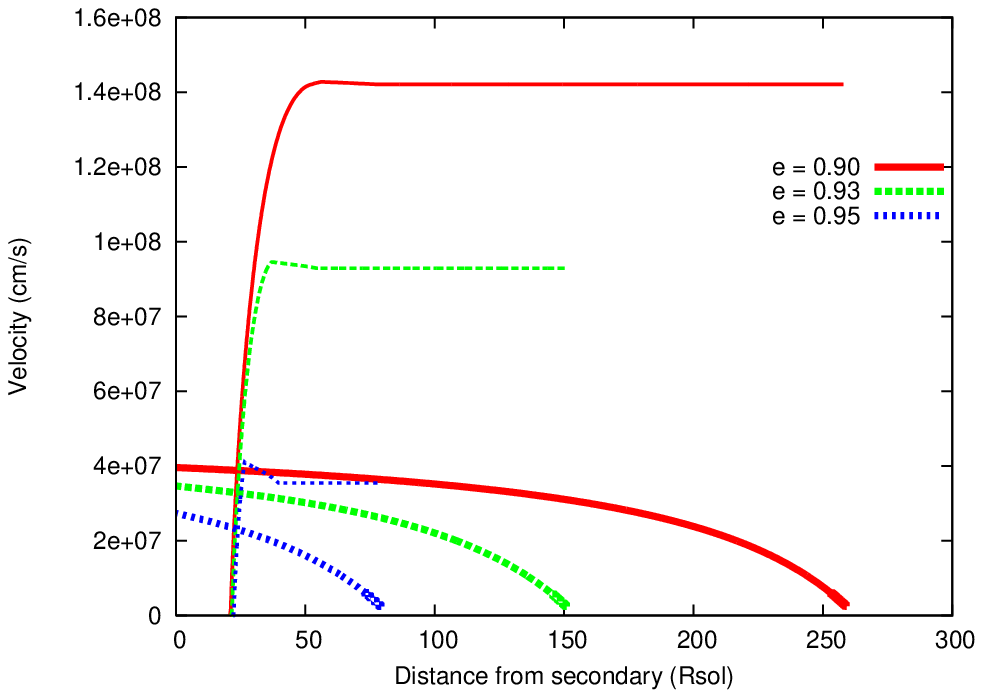}}} \\
      \resizebox{80mm}{!}{{\includegraphics{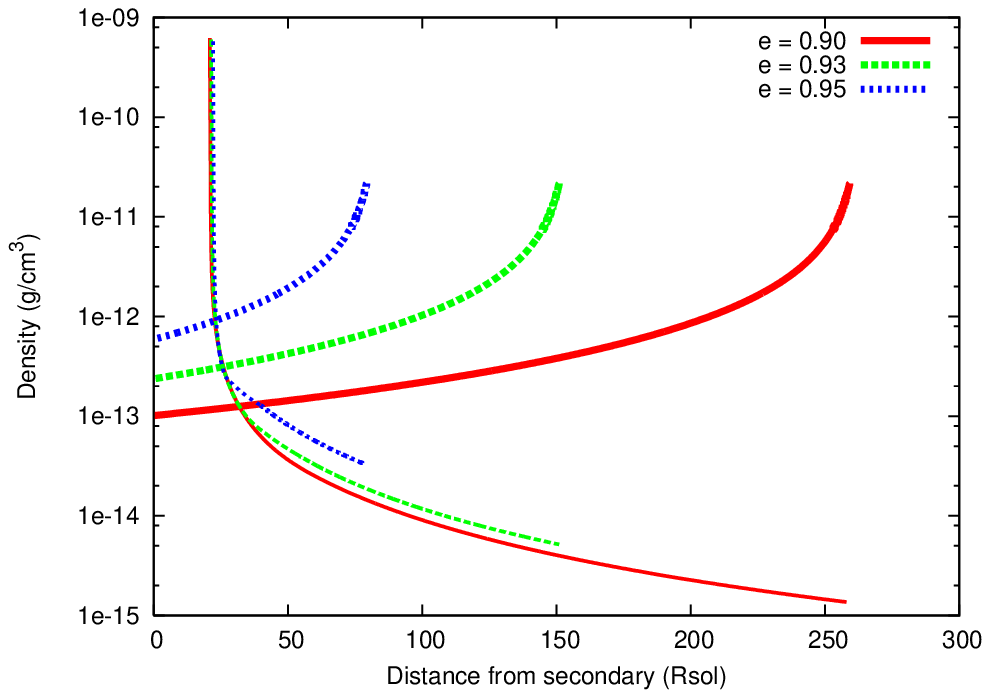}}} \\
    \end{tabular} 
\caption[]{The ram pressure (top), velocity (middle), and density
(bottom) of the radiatively driven winds as a function of distance
from the companion star and orbital eccentricity. The separation of
the stars corresponds to periastron in each instance. In each plot the
companion's wind (thin lines) starts at 20 \Rsol. Radiative inhibition
of the companion's wind by the primary's radiation field is included,
where coupling $C_2$ is used. The top panel shows whether a stable
momentum balance can be achieved \citep[indicated when an intersection
occurs with opposing gradients, see Fig.~2 of][]{Stevens:1992}. The
braking radius (Fig.~\ref{fig:radbrake}) for each scenario is
indicated by the marks at the top of the uppermost plot. No stable
balance exists for $e\gtsimm 0.95$ unless braking is important. The
parameters used in the line driving calculations are noted in
Table~\ref{tab:cakparameters}.}
\label{fig:radinh_ramp}
\end{center}
\end{figure}

\subsection{Collapse of the wind collision region onto the secondary star}
\label{subsec:WCRcollapse}

Examining the ram pressure balance between the reduced velocity winds
discussed in the previous section reveals that, when $e=0.9$, the
equilibrium is stable when the companion's $k$ and $\alpha$
(i.e. coupling $C_2$) are used in the radiative inhibition
calculations (Fig.~\ref{fig:radinh_ramp}). An increased orbital
eccentricity reduces the ram pressure of the companion's wind at
periastron, due to the increasing influence of the primary's radiation
field as the stellar separation is reduced. At $e = 0.95$ there is no
stable balance between the winds at periastron, which is indicated by
the lack of an intersection between the ram pressure curves for the
winds with opposing gradients. If coupling $C_1$ is used to perform
these calculations there is no stable balance point at periastron even
when $e = 0.90$. Clumping within the winds, a temporary increase in
the primary mass-loss rate, and the gravitational influence of the
companion star may further reduce the likelihood of a stable balance
between the winds. Referring back to Fig.~\ref{fig:radbrake} it is
clear that in the scenarios where a collapse is predicted, subsequent
radiative braking of the primary star's wind before it reaches the
companion star's surface should also occur. This is of course a very
complicated picture, and in reality it is difficult to predict the sum
of all of the different mechanisms described above without a single
model which self-consistently includes all of the appropriate physics
which we have shown is important. This is beyond the scope of the
current work, but will be addressed in future models.

Is there any observational evidence for a collapse of the WCR onto the
companion star? Let us first consider the X-ray emission. While
radiative inhibition can dramatically reduce the hard X-ray emission
around periastron (see Fig.~\ref{fig:radinh_specs}), it cannot by
itself explain the long duration of the X-ray minimum and its slow
recovery, since the stars separate very quickly after
periastron. Furthermore, the preshock wind speed of the companion
inferred from the fits to the \textit{XMM-Newton} data remains roughly
constant during the X-ray minimum ($\phi\simeq1.0-1.03$), whereas the
inhibition calculations indicate that it should rapidly rise
(Fig.~\ref{fig:radinh_psvels}).

However, if the WCR collapses onto the surface of the companion star
it is unclear how quickly the companion wind will be able to
re-establish itself. It is entirely possible that once the primary
wind collides directly with the surface of the companion star its
continued ram pressure will prevent the companion star from developing
a wind towards the primary star until long after periastron has passed
\citep[see the hydrodynamic simulations of such an eventuality in
][]{Pittard:1998b}. Indeed, the asymmetry of the X-ray minimum about
periastron passage indicates that the WCR in \etacar exhibits
hysterisis in this sense.

\begin{figure}
\begin{center}
    \begin{tabular}{l}
      \resizebox{80mm}{!}{{\includegraphics{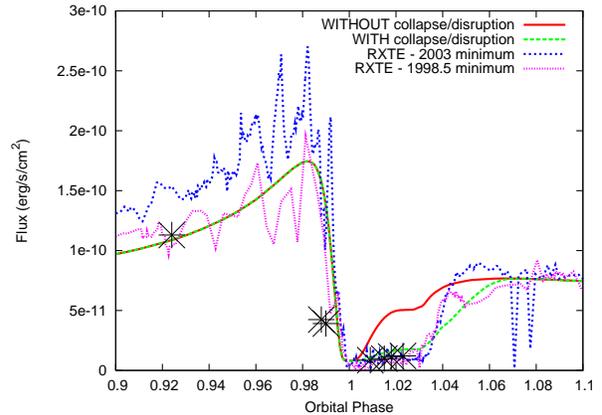}}} \\
    \end{tabular} 
\caption[]{The 2-10 keV X-ray lightcurve showing the effect of a
collapse of the WCR onto the companion star between phases 1.0 and
1.03. During this time the only emission arises from further
downstream. The other parameters of the model are the same as those
for the $\eta = 0.18$ lightcurve shown in
Fig.~\ref{fig:results_lc_los}. The crosses represent the integrated
fluxes from the individual models with lower companion wind velocities
discussed in \S~\ref{subsec:rad_braking}.}
\label{fig:wcrcollapse}
\end{center}
\end{figure}

To further examine this possibility a mock collapse of the WCR has
been incorporated into the dynamic model in \S~\ref{sec:results}. The
collapse is assumed to begin at $\phi = 1.0 $ and to end at $\phi =
1.03$. The emission from the region of the WCR which is affected by
the collapse is set to zero, while shocked gas further downstream (and
upstream after the recovery) emits as normal. The simulated observed
emission following this addition to the model is displayed in
Figs.~\ref{fig:results_hr}, ~\ref{fig:results_specs} and
\ref{fig:wcrcollapse}. The reduction in the hard band emission
relative to the previous models is clearly evident in the spectra
shown in Fig.~\ref{fig:results_specs}, and a much better match to the
observations during the minimum compared to the `standard' model is
obtained. The fit to the lightcurve minimum is also considerably
improved (Fig.~\ref{fig:wcrcollapse}). While there are remaining
differences in the hardness ratio (Fig.~\ref{fig:results_hr}), the
similarity of its shape during the X-ray minimum is intriguing
(specifically the slower change from softer to harder emission in the
second half of the minimum). Referring back to our conclusion in
\S~\ref{subsubsec:los} that lines-of-sight which favour the companion
star in front at periastron are excluded still remains true in the
collapse scenario as the pre/post-minimum flux level provides a poor
match to the observed lightcurve.

Therefore, these results provide significant support for a collapse of
the WCR onto the companion star around periastron.

During the collapse of the WCR the companion star may accrete some of
the primary star wind \citep{Soker:2005}. Taking the mass accretion
rate to be $\dot{M}_{\rm acc}\simeq \pi\rho_{1}(r)R_{\rm
acc}^{2}v_{1}(r)$ \citep[e.g. Eq.(15) in][]{Akashi:2006}, where
$\rho_{1}(r)$ and $v_{1}(r)$ are the density and velocity profiles
calculated for the primary star wind and the accretion
radius\footnote{In \cite{Akashi:2006} the accretion radius is
determined under the assumption that the companion star is not driving
a wind in any direction. In contrast, we assume that the companion
star drives a wind on the side facing away from the primary star,
hence accretion only occurs for gas striking the near-side of the
companion star and the accretion area becomes $\sim \pi R_{2} ^{2}$}
$R_{\rm acc} = R_{2}$ (Table ~\ref{tab:testparameters}). We find the
total accreted mass during the collapse to be $\sim 7\times10^{-8}
\Msol$, which is approximately one to two orders of magnitude lower
than values noted by \cite{Kashi:2008}.

\section{Conclusions}
\label{sec:conclusions}

We have constructed the first 3D model of the colliding winds in
\etacar with spatially extended, energy dependent, X-ray
emission. Coriolis forces skew the WCR and wrap the far downstream
regions around the stars. Attenuation of the X-rays through the
unshocked winds of both stars, and through the cold, dense layer of
postshock primary wind is considered. The simulations place
constraints on the orbital orientation and nature of \etacar through
comparison with observations from the \textit{RXTE} and
\textit{XMM-Newton} satellites. The main findings from this work can
be summarized thus:
\begin{itemize}
\item 3D effects, specifically the skew to the WCR due to orbital
   motion, lead to an asymmetry of the X-ray luminosity either side of
   the minimum. Circumstellar absorption is usually dominated by the
   un-shocked stellar winds. For a very brief period ($\delta \phi
   \sim 0.01$) the absorption may be dominated by material in the cold
   postshock primary wind.

\item Inclination angles similar to the polar axis angle of the
   Homunculus nebula \citep{Smith:2006} are favoured.

\item Orientations with the secondary star in front of the primary
   around periastron passage are supported. The model results favour a
   line-of-sight angled at $0 - 30^{\circ}$ to the semi-major axis in
   the prograde direction, in agreement with \cite{Nielsen:2007} and
   \cite{Okazaki:2008}. Orientations with the secondary star in front
   of the primary or at quadrature during periastron passage are
   excluded.

\item With terminal wind velocities fixed at $v_{\infty
   1}=500\thinspace \rm{km\thinspace s}^{-1}$ and $v_{\infty
   2}=3000\thinspace \rm{km\thinspace s}^{-1}$ for \etacar and the
   companion star respectively (our `standard' model), the best fit to
   the 2-10 keV \textit{RXTE} lightcurve was obtained with mass-loss
   rates of $\dot{M}_1 = 4.7\times10^{-4}\Msolpyr$ and $\dot{M}_2 =
   1.4\times10^{-5}\Msolpyr$. These values are similar to an earlier
   determination by PC02 though the mass-loss rates are slightly
   higher and the wind momentum ratio, $\eta$ is slightly lower. This
   difference reflects changes in the adopted distance and
   interstellar absorption to the system, and to difficulties in
   separating other, more spatially extended, components from the
   large \textit{RXTE} beam. It may also reflect the lack of mixing at
   the CD in the dynamic model used in this work, compared to the
   hydrodynamical simulations of PC02.

\item Significant discrepancies between the observed and `standard'
   model lightcurves and spectra exist through the X-ray minimum
   in. We conclude that a wind eclipse is not the sole cause of the
   minimum, in contrast to the recent conclusions of
   \cite{Okazaki:2008} who modelled the X-ray emission as a
   mono-energetic point source and considered only the lightcurve.

\item An examination of the importance of radiative inhibition has
   revealed that around periastron the preshock velocity of the
   companion wind is likely to be significantly reduced. The secondary
   wind is then shocked to much lower temperatures and cools
   rapidly. As a result the hard X-ray flux is efficiently
   quenched. All future models must properly account for the effect of
   radiative inhibition.

\item While the hard X-ray flux can be significantly reduced during
   periastron passage, the long duration of the minimum after
   periastron when the stars are rapidly separating hints that
   something very extraordinary happens. We believe that the most
   likely explanation is the collapse of the WCR onto the companion
   star, perhaps helped by a short increase in the mass-loss rate of
   the primary wind. Incorporating a mock collapse into the model
   between phases 1.0 and 1.03 yields significantly better agreement
   between the model and data. 

\end{itemize}

The dynamical model described in this paper provides insight into the
role of 3D effects in highly eccentric binaries. Its failings to
produce synthetic X-ray emission which is a good match to the observed
data have highlighted the importance of the stellar radiation fields
on the dynamics of the winds and the structure and cooling of their
collision region. To advance this work further requires 3D
hydrodynamical models which incorporate cooling, gravity and the
driving of the stellar winds.

\subsection*{Acknowledgements}
ERP thanks the University of Leeds for funding. JMP gratefully
acknowledges funding from the Royal Society. We acknowledge the help
of the \textit{RXTE} Guest Observer Facility for supporting this
work. This research has made use of NASA's Astrophysics Data
System. This research has made use of data obtained from the High
Energy Astrophysics Science Archive Research Centre (HEASARC) provided
by NASA's Goddard Space Flight Center.

\bibliography{parkin}{}

\label{lastpage}


\end{document}